\documentclass[9pt,twocolumn,twoside]{pnas-new}
\usepackage{graphicx}
\usepackage{subfig} %To put subfig caption at top use [position=top]
\usepackage{braket}

\setboolean{displaywatermark}{false}
\templatetype{pnasresearcharticle} % Choose template
\newcommand{\refcite}[1]{Ref.~\cite{#1}}

\renewcommand{\approx}{\simeq}

\newcommand{\ce}{\mathcal{E}}

\definecolor{wrongultramarine}{rgb}{1,0.5,0}

\newcommand{\rd}{{\rm d}}

\newcommand{\beq}{\begin{equation}}
\newcommand{\eeq}{\end{equation}}
\def\bea{\begin{eqnarray}}
\def\eea{\end{eqnarray}}

\usepackage{pdfpages}
\usepackage{pgffor}
\makeatletter
\AtBeginDocument{\let\LS@rot\@undefined}
\makeatother

\title{Resonant thermal Hall effect of phonons coupled to dynamical defects}

\author[a]{Haoyu Guo}
\author[a]{Darshan G. Joshi}
\author[a,b]{Subir Sachdev}

\affil[a]{Department of Physics, Harvard University, Cambridge MA 02138, USA}
\affil[b]{School of Natural Sciences, Institute for Advanced Study, Princeton, NJ-08540, USA}
\leadauthor{Guo}

\significancestatement{Modern quantum materials display numerous phases of electronic matter with many-particle quantum entanglement between the electrons. However, this entanglement is notoriously difficult to characterize experimentally. Recent experiments have shown that thermal Hall effect (when in a magnetic field, there is heat flow in a direction transverse to a temperature gradient) is a sensitive probe of the many-electron quantum state. We propose that these observations detect the scattering of lattice vibrations (phonons) from electronic impurities, and compute the influence of the electronic dynamics on the heat carried by the phonons. We also propose a specific mechanism for thermal Hall effect in the `pseudogap' state of the cuprates, the entangled state which leads to high temperature superconductivity at smaller electron density.}

\authorcontributions{Please provide details of author contributions here.}
\authordeclaration{Please declare any competing interests here.}
\correspondingauthor{\textsuperscript{2}To whom correspondence should be addressed. E-mails: haoyuguo@g.harvard.edu, sachdev@g.harvard.edu}


\begin{abstract}
We present computations of the thermal Hall coefficient of phonons scattering off a defect with multiple energy levels. Using a microscopic formulation based on the Kubo formula, we find that the leading contribution perturbative in the phonon-defect coupling is proportional to the phonon lifetime, and has a
`side-jump' interpretation. Consequently, the thermal Hall angle is independent of the phonon lifetime. The contribution to the thermal Hall coefficient is at resonance when the phonon energy equals a defect level spacing. Our results are obtained for three different defect models, which apply to different correlated electron materials. For the pseudogap regime of the cuprates, we propose a model of phonons coupled to an impurity quantum spin in the presence of quasi-static magnetic order with an isotropic Zeeman coupling to the applied field, and without spin-orbit interaction.
\end{abstract}

\dates{This manuscript was compiled on \today}
\doi{\url{www.pnas.org/cgi/doi/10.1073/pnas.XXXXXXXXXX}}

\begin{document}

\maketitle
\ifthenelse{\boolean{shortarticle}}{\ifthenelse{\boolean{singlecolumn}}{\abscontentformatted}{\abscontent}}{}

% If your first paragraph (i.e. with the \dropcap) contains a list environment (quote, quotation, theorem, definition, enumerate, itemize...), the line after the list may have some extra indentation. If this is the case, add \parshape=0 to the end of the list environment.
\dropcap{T}he thermal Hall effect has recently emerged as a powerful probe of correlated electron materials. The thermal Hall conductivity, $\kappa_H$, is the analog of the electrical Hall conductivity, with heat currents and temperature differences replacing electrical currents and voltages.
Recent experiments in the cuprates \cite{Grissonnanche2019,Grissionnanche2020,Boulanger2020,boulanger2021}, magnetic insulators \cite{chen2021large}, metallic spin ice $\rm{Pr}_2\rm{Ir}_2\rm{O}_7$\cite{Machida2022}, ferroelectric $\rm{Sr}\rm{Ti}\rm{O}_3$ \cite{Li2020}, and spin liquid candidate $\rm{RuCl}_3$ \cite{TailleferRuCl} have observed a large $\kappa_H$, but various experimental characteristics support an interpretation in which the energy currents are primarily carried by the phonons, rather than the correlated electrons.
At low temperatures, the contribution of gapless acoustic phonons should dominate $\kappa_H$.
Similar to the charge Hall effect of electrons \cite{RMP_AHE}, acoustic phonons can contribute to $\kappa_H$ through an intrinsic effect arising from the Berry curvature \cite{Shi12} of phonon bands, but this effect is too small to account for the observations \cite{kivelson20,Barkeshli12,ye2021phonon,ZhangTeng21}.
Attention therefore turned to the influence of static impurities on the phonon transport of heat \cite{kivelson20,Guo2021, Flebus21}: while larger than the intrinsic contribution because it is enhanced by the phonon mean free path, this extrinsic contribution is also insufficient to provide a unified explanation of the data, as we discuss below.

 In this paper, we provide a theory of phononic $\kappa_H$ which addresses two sets of questions. (A) How does the phonon sense the magnetic field, {\it i.e.\/} how is time-reversal symmetry broken for phonons? (B) How does a time-reversal symmetry breaking in the phonon sector translate to a thermal Hall effect? Before discussing our answers to these questions, it is helpful to review the experimental observations in the cuprates \cite{Grissonnanche2019,Grissionnanche2020,Boulanger2020,boulanger2021}, which our theory tries to explain. (1) In the hole-doped cuprates, the phononic thermal Hall effect appears immediately upon decreasing doping from the Fermi Liquid to the pseudogap regime. (2) The phonon contribution to $\kappa_H$ continues into the antiferromagnetic insulator phase without visible features when crossing the Neel transition. (3) The thermal Hall effect is quasi-isotropic, meaning that $\kappa_{xy}$ (measured at $\vec{B}\parallel\hat{z}$) and $\kappa_{yz}$ (measured at $\vec{B}\parallel\hat{x}$) have similar orders of magnitude and temperature ($T$) dependence. (4) The Hall angle $|\Theta_H|\equiv |\kappa_H|/\kappa_{xx}$ remains in the range $0.002$-$0.006$ across a wide range of hole-doped cuprates and mother compounds with large variations in $\kappa_{xx}$ and sample quality.

 Previous attempts \cite{kivelson20,ye2021phonon,ZhangTeng21,Guo2021} to answer question (A) have employed an effective field theory mindset: we consider the leading order time-reversal breaking term for phonons, called the phonon Hall viscosity (PHV) \cite{Barkeshli12}. As the phonon thermal Hall effect appears also in an insulator, it is believed that PHV comes primarily from spin-lattice coupling. However, it is shown in \cite{ye2021phonon} that PHV is proportional to spin-orbit coupling, which is weak in the cuprates, and therefore unable to explain the experiments. The failure of the PHV approach calls for the inclusion of additional degrees of freedom.

It is now useful to recall a historical puzzle in the  longitudinal thermal conductivity of glasses, where phonon scattering was observed to be anomalously large at low temperature ($T$). It was proposed \cite{AHV,WAP} that the scattering of phonons off two-level systems, {\it i.e.\/} dynamical defects, could provide the needed enhancement, and this explanation has been since consistent with observations \cite{Krivchikov}. In our work, we address question (A) by examining the role of dynamical defects in heat transport in correlated electron systems. We examine processes in which phonons resonant with the level splitting of the defect are absorbed and re-emitted, as in Refs.~\cite{AHV,WAP}. The phonon-defect coupling (which can have forms labeled A, B, C, appropriate to different physical situations) endows this process with a chiral character, leading to a resonant enhancement of the thermal Hall effect.
We note that Sun {\it et al.\/} \cite{sun2021} have also recently made a related proposal, although their defect model and computational method are different from ours. We will argue that model B, describing the two Zeeman-split levels of an impurity in an antiferromagnetic environment, provides an attractive description of observations in the cuprate pseudogap metal \cite{Grissonnanche2019,Grissionnanche2020,Boulanger2020,boulanger2021}. Our models also connect to recent observations \cite{Machida2022} in a metallic spin ice compound.
We emphasize that the three models described below apply to \emph{different} microscopic situations. However, their leading phonon thermal-Hall effects are universally described by the side-jump mechanism, which we discuss below.
%Therefore, it is not meaningful to compare $\kappa_{H}$ of different models.

As for question (B), the possibilities that a quasiparticle can contribute to Hall effects have been enumerated in the literature of charge Hall effects \cite{RMP_AHE,Sinitsyn06,Sinova07}: Berry curvature, skew scattering and side jump. In the context of phonon thermal Hall effect, the Berry curvature contribution is weak because it is independent of scattering, and hence not enhanced by the large phonon mean-free path. For extrinsic contributions due to scattering, most works \cite{Mori2014,kivelson20,Guo2021, Flebus21, sun2021} have focused on the skew-scattering mechanism.  With a phonon skew scattering time $\tau_{\rm skew}$, and a total mean-free time $\tau_{\rm ph}$, we can write the longitudinal ($\kappa$) and Hall conductivities as
\begin{equation}
    \kappa_{xx} \sim \frac{1}{3} C_v v^2 \tau_{\rm ph} \quad, \quad |\kappa_H| \sim \frac{1}{3} C_v v^2 \tau_{\rm ph}^2 \tau_{\rm skew}^{-1}\,,
\end{equation}
where $C_v$ is the specific heat, and $v$ is a typical acoustic phonon velocity. The above estimates are valid when $\tau_{\rm skew}^{-1}\ll\tau_{rm ph}^{-1}$, which holds for the linear response to external magnetic field. This contribution is unfavorable in explaining the cuprate experiments for two reasons: First, the Hall angle $|\Theta_H| = |\kappa_H|/\kappa_{xx} \sim \tau_{\rm ph}/\tau_{\rm skew}$ depends on the ratio between the two scattering time scales. As mentioned before, the values of $|\Theta_H|$ lie in the range 0.002-0.006 for hole-doped cuprates despite large variations in $\kappa_{xx}$ and sample quality, and this is unlikely for skew scattering because $\tau_{\rm ph}$ and $\tau_{\rm skew}$ can have distinct microscopic origins. Second, the skew scattering contribution has a parity problem \cite{Guo2021,sun2021,Mori2014} which causes it to vanish when there is only one scattering channel that respects inversion symmetry of phonon wave function. A nonzero skew scattering $\kappa_H$ typically requires interference between multiple scattering channels and hence depend on material specific details. Therefore, our work will focus on the side-jump mechanism.
We will show that the side-jump contribution, in contrast, has $\kappa_H \sim \tau_{\rm ph}$, and so $\Theta_H$ is naturally independent of the phonon lifetime. We estimate the resulting value of $\Theta_H$ in Section~2.\ref{ThetaH} and find it consistent with observations \cite{boulanger2021}. Furthermore, there is no parity problem for the side jump mechanism, and $\kappa_H$ is nonzero for a generic scattering channel.

The computation of $\kappa_H$ in interacting systems is challenging because we need to account for the energy of the interactions, and subtract the non-transport heat current due to the `energy magnetization' \cite{CHR,QinPRL,Kapustin2021,Shi12}. In our model, we have the additional complexity of keeping
track of the energy of the dynamical defect. We use the recently developed method of Kapustin and Spodyneiko \cite{Kapustin2021} to overcome these difficulties using a Kubo formula, and present a systematic expansion of the defect contribution to $\kappa_H$ in powers of the phonon-defect coupling. Our main results are obtained at second order, and are the analog of the `side-jump' contributions to the electrical Hall effect \cite{RMP_AHE,Sinitsyn06,Sinova07}. The analog of the `skew-scattering' contribution (considered in Ref.~\cite{sun2021}) appears at fourth order, among many other contributions.

% The previous works \cite{kivelson20,Guo2021, Flebus21} on phonons coupled to static impurities considered the skew scattering contribution. With a phonon skew scattering time $\tau_{\rm skew}$, and a much shorter non-skew scattering time $\tau_{\rm ph}$, we can write the longitudinal ($\kappa$) and Hall conductivities as
% \begin{equation}
%     \kappa_{xx} \sim \frac{1}{3} C_v v^2 \tau_{\rm ph} \quad, \quad |\kappa_H| \sim \frac{1}{3} C_v v^2 \tau_{\rm ph}^2 \tau_{\rm skew}^{-1}\,,
% \end{equation}
% where $C_v$ is the specific heat, and $v$ is a typical acoustic phonon velocity. Then we have for the Hall angle $|\Theta_H| = |\kappa_H|/\kappa_{xx} \sim \tau_{\rm ph}/\tau_{\rm skew}$. The values of $|\Theta_H|$ have been observed \cite{boulanger2021} to be in the range 0.002-0.006 across a wide range of hole-doped cuprates with large variations in $\kappa_{xx}$
% and sample quality, and this is unlikely to realized by skew scattering. In contrast, we will show that the side jump contribution has $\kappa_H \sim \tau_{\rm ph}$, and so $\Theta_H$ is independent of the phonon lifetime. We estimate the resulting value of $\Theta_H$ in Section~2.\ref{ThetaH} and find it consistent with observations \cite{boulanger2021}.

\section{Models and computation}

Before precisely specifying our models A, B, C, we highlight their main properties.

\begin{figure}
  \centering
  \includegraphics[width=0.85\columnwidth]{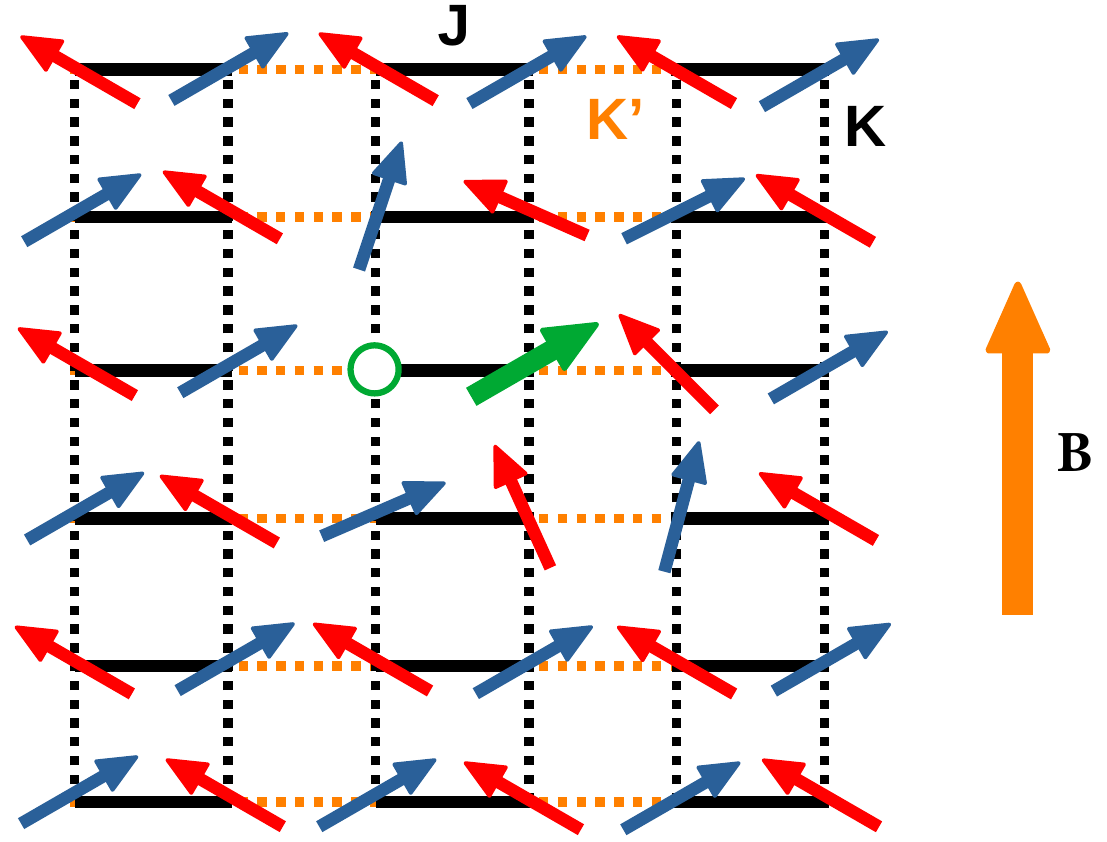}
  \caption{Illustration of model B. The defect spin (green) is polarized by local quasi-static magnetic order. When the external field $B=0$, the nearby spins are coplanar and there is no thermal Hall response. External field $B$ induces additional canting to the spins which results in nonzero $\kappa_H$.}\label{fig:mag}
\end{figure}

In models A and B, we consider a spin-1/2 defect embedded in an environment with local magnetic order (see Fig.~\ref{fig:mag}). Experiments have shown \cite{Grissonnanche2019, Grissionnanche2020,Boulanger2020,chen2021large} that magnons are not important for the thermal Hall effect, so we can treat the nearby spins as frozen.
The primary coupling of the external magnetic field is via  the Zeeman term, and so is independent of the orientation relative to the lattice. In the simpler model A, the coupling between the spin and phonons arises from the spin-orbit interactions, and
depends upon the precise manner in which lattice and time-reversal symmetries are broken near the defect (as in the Rashba term on surfaces \cite{Duine15}).
The spin-orbit interaction is not required for model B, provided the magnetic order has a suitable non-coplanar structure near the impurity. The local non-coplanar order may be related to the spin glass behavior recently observed in pseudogap cuprates \cite{Julien19}.

In model C, we consider a defect which consists of a ground state singlet ($\ell=0$), and an excited triplet ($\ell=1$) whose degeneracy is split by an orbital coupling to an external field (and so is strongly dependent upon the relative orientation of the field and the lattice). This model is similar to the three-level model considered by Sun {\it et al.\/} \cite{sun2021}.

We consider 3-dimensional phonons coupled to defects with Hamiltonian
\begin{equation}\label{}
       H=H_{\rm ph}+H_{\rm def}+H_{\rm ph-def}\,.
\end{equation}
The phonon Hamiltonian includes a dissipation term
     \begin{equation}\label{eq:Hph}
       H_{\rm ph}=\sum_{p} \frac{\pi_p^i \pi_p^i}{2m}+\frac{1}{2}\sum_{pq}u_p^i C_{pq}^{ij}u^j_q+H_{\rm dis}\,,
     \end{equation}
We use $i,j,k,\dots=x,y,z$ to denote Cartesian indices and they are subject to Einstein summation convention, and we use $p,q,r,\dots$ to denote site indices, and they are not implicitly summed. We use $u$ to denote the lattice displacement of an ion, and $\pi$ to denote its momentum, and they satisfy the commutation relation
$
       [u_p^i,\pi_q^j]=i\delta_{pq}\delta^{ij}\,.
$
In \eqref{eq:Hph}, $m$ is the ion mass and $C_{pq}^{ij}$ is the elastic coupling between neighboring ions. The elastic coupling is chosen such that in the continuum limit it describes isotropic phonons with longitudinal velocity $c_L$ and transverse velocity $c_T$. The dissipation in $H_{\rm dis}$, leads to a lifetime $\tau_{\rm ph} = 1/\Gamma_{\rm ph}$ in the phonon Green's function. We assume phonons remain well-defined quasiparticles, so $\Gamma_{\rm ph}\ll T$. The phonon-defect coupling is treated perturbatively because of the heavy ion mass $m$.
We work in units $k_B=\hbar=1$.
%The phonon-defect coupling $H_{\rm ph-def}$ might shift the equilibrium positions of the ions, but this doesn't affect the thermal Hall effect because we will only need retarded or advanced Green's functions which are independent of equilibrium positions.

\begin{figure}
  \centering
  \subfloat[]{\includegraphics[width=0.2\textwidth]{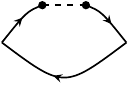}} ~~~~
  \subfloat[]{\includegraphics[width=0.2\textwidth]{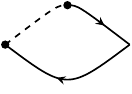}}
% \begin{subfigure}[t]{0.45\columnwidth}
% \centering
% \includegraphics[width=0.5\textwidth]{Fig2a.pdf}
%   \caption{}\label{fig:diagramsa}
% \end{subfigure}
% \begin{subfigure}[t]{0.45\columnwidth}
% \centering
% \includegraphics[width=0.5\textwidth]{Fig2b.pdf}
%   \caption{}\label{fig:diagramsb}
% \end{subfigure}
\caption{Feynman diagrams contributing to the side-jump thermal Hall effect. The left and the right ends are energy current vertices. The solid line is phonon propagator, and the dashed line is defect propagator. The black circle is phonon-defect coupling. The arrows label momenta of phonons. The contribution of diagram (a) to the energy current-energy current correlation function is given by Eq. (S62) (Model C) and (S104) (Models A and B), and diagram (b)'s contribution is given by Eq. (S63) (Model C) and Eq. (S105) (Models A and B). }\label{fig:diagrams}
\end{figure}

Although the Boltzmann equation is not directly applicable, it provides guidance to the $\Gamma_{\rm ph}$ dependence of the expansion in the phonon-defect coupling $H_{\rm ph-def}$.
In the electrical Hall effect \cite{Sinova07}, a Feynman diagram corresponds to intrinsic, side-jump and skew scattering when 0,1,2 of its two energy current vertices are intra-band respectively. In the DC limit, an intra-band vertex function is attached to a pair of retarded and advanced Green's function with nearly identical denominator, and so contributes a large factor of $1/\Gamma_{\rm ph}$. Consequently, the above classification also counts the degree of divergence in the $\Gamma_{\rm ph}\to 0$ limit. To second order in $H_{\rm ph-def}$,
phonons have no skew-scattering (because there is no 2-phonon-irreducible diagram), and we will compute the most singular term due to the side-jump diagrams in Fig.~\ref{fig:diagrams}, proportional to $1/\Gamma_{\rm ph}$. Fig.~\ref{fig:diagrams}a describes the inter-band coherence of phonons induced by $H_{\rm ph-def}$, similar to charge Hall effect \cite{Sinitsyn06}.
$H_{\rm ph-def}$ will also contribute to the energy current vertex, and this leads to Fig.~\ref{fig:diagrams}b, which captures the quantum coherence between the phonon and the defect wavefunction, unique to a single-phonon process. Semiclassically, this effect implies that the absorption or emission of a phonon does not happen exactly at the position of the defect, but the coordinate is shifted similar to the coordinate shift in the charge Hall effect.
%%%%%%%%%%%%%%%%%%%%%%%%%%%%%%%%%%%%%%%%%%%%%%%%%%%%%%%%%%%%
%%%%%%%%%%%%%%%%%%%%%%%%%%%%%%%%%%%%%%%%%%%%%%%%%%%%%%%%%%%%%%%%%

\subsection{Lattice formalism for thermal Hall effect}

In this subsection we briefly review the lattice formalism derived by Kapustin and Spodyneiko \cite{Kapustin2021}. We assume the Hamiltonian can be decomposed into a sum of local terms $H=\sum_p H_p$. Here $H_p$ is localized around site $p$ and is Hermitian. For two sites $p,q$ sufficiently far apart, $[H_p,H_q]=0$. On the lattice, the energy current is not a vector field, but a two-point operator denoted as $J^E_{pq}$, measuring the energy flowing from site $q$ to site $p$. Using Heisenberg's equation, we find $J^E_{pq}=-i[H_p,H_q]$. The total current flowing across the plane $x=a$ is then
\begin{equation}\label{}
      J^E(\delta f)=\frac{1}{2}\sum_{pq}J^E_{pq}\left(f(q)-f(p)\right)\,,
\end{equation}
where $f(p)=\theta(a-x(p))$ is the step function at $x=a$. We mention in passing that the two-point operator $J^E_{pq}$ is a 2-chain, the 1-point function $f(p)$ is a 1-cochain and $\delta f(p,q)=f(q)-f(p)$ is a 2-cochain and also the exterior derivative of $f$.

The energy current is driven by Luttinger's fictitious gravitational field, which is turned on adiabatically
$ \Delta H(t)=\epsilon e^{st} \sum_{p}g(p)H_p$ for $t\leq 0$, and $\Delta H(t>0)=\Delta H(t=0)$. The switching rate $s>0$ is infinitesimal. The function $g(p)=-y(p)/L$ describes a uniform gravitational field with unit potential drop across the sample. The derivative of the thermal conductance with respect to arbitrary parameter in the Hamiltonian
is \cite{Kapustin2021}
\begin{align}
      \rd \kappa(f,g)=&\rd \left[\beta^2 \lim_{s\to 0^+} \int \rd t e^{-st}\Braket{\Braket{J^E(\delta f,t);J^E(\delta g)}}\right] \nonumber \\
      &-2\beta \mu^E(\delta f\cup \delta g)\,. \label{eq:Conductance}
\end{align}
The first line is the usual Kubo formula and $\Braket{\Braket{\cdot;\cdot}}$ refers to the Kubo canonical pairing: $\Braket{\Braket{A;B}}=\beta^{-1}\int_0^\beta \rd \tau \Braket{A(-i\tau)B(0)}$. The second line is the magnetization correction, which is explicitly:
\begin{align}
       &\mu^E(\delta f\cup \delta g)  =\frac{1}{6} \sum_{pqr}\mu^E_{pqr} (f(q)-f(p))(g(r)-g(q))\,, \nonumber \\
       & \mu^E_{pqr}  =-\beta\left[\Braket{\Braket{\rd H_p;J^E_{qr}}}+\Braket{\Braket{\rd H_r;J^E_{pq}}} \right. \nonumber \\
       &~~~~~~~~~\left. +\Braket{\Braket{\rd H_q;J^E_{rp}}}\right]\,. \label{eq:muE_main}
\end{align}
The thermal Hall conductivity can be extracted from \eqref{eq:Conductance} by averaging the current $J^E(\delta f)$ over the sample (this is equivalent to setting $f(p)=-x(p)/L$), and also antisymmetrizing with respect to $f,g$. Therefore the final formula for the thermal Hall conductivity is
\begin{equation}\label{eq:kappaH_main}
      \rd \kappa_{H}=\frac{1}{2L^{d-2}}\left(\rd\kappa(f,g)-\rd\kappa(g,f)\right)\,,
\end{equation}
where $f(p)=-x(p)/L$, $g(p)=-y(p)/L$ and $L^{d-2}$ is the cross section area of the sample divided by sample length. As a sanity check, we show later in  supplement \cite{Supp} that for a non-interacting phonon Hamiltonian,  \eqref{eq:kappaH_main} agrees with the Berry curvature formula in \refcite{Shi12}.

To evaluate the thermal Hall conductivity, we need to specify the integration path in the parameter space. We will  assume weak phonon-defect coupling and perturb in the coupling constant: it is natural to choose this as the differentiation parameter for it is easy to integrate, and the thermal Hall effect vanishes in its absence.

Further details of the computation of $\kappa_H$ appear in the supplement  \cite{Supp}. In particular, we show that the energy magnetization correction is subdominant in power of $1/\Gamma_{\rm ph}$ because the phonon Green's functions attached to the vertices are either all retarded or all advanced. The computation yields the following semiclassical formula for thermal Hall effect:
\begin{equation}\label{eq:kappaH_semiclassical}
\begin{split}
  \kappa_H=&\frac{1}{2}\sum_{a}\int \frac{\rd^3 k}{(2\pi)^3} \frac{(- n_B'(\ce_a(k))}{T\Gamma_a(k)}j^E_{\text{on-shell},x}j^E_{\text{side-jump},y}\\
  &-(x\leftrightarrow y)\,.
\end{split}
\end{equation}
Here $k$ denotes phonon momentum, $a=1\dots 6$ runs over positive and negative frequency modes of the three phonon bands, and $n_B$ is the Bose function. $\ce_a(k)$ is the phonon energy and $1/\Gamma_a(k)$ is the phonon lifetime. $j^E_\text{on-shell}$ and $j^E_\text{side-jump}$ are two types of energy currents. $j^E_{\text{on-shell}}$ is the usual energy current $j^E_{\text{oh-shell},x}=\ce_a(k)v_{a,x}(k)$ where $v_{a}(k)$ is the bare velocity of phonon mode. The side-jump energy current is $j^E_{\text{side-jump},y}=\ce_a(k) v_{a,y}^{\rm (sj)}(k)$ where the side-jump velocity $v_a^{\rm (sj)}$ contains two parts
\begin{equation}v_{a,y}^{\rm (sj)}=v_{aa,y}^{\rm (sj)}+\sum_{b:\ce_b\neq \ce_a} v_{ba,y}^{\rm (sj)}\,.\end{equation}
The first term $v_{aa}^{\rm sj}$ (Eq.(S129)) comes from the momentum dependence of phonon-defect coupling, whose expression depends on particular form of the coupling. The physical meaning of this term is the renormalization of intraband velocity due to interactions. For the three models we consider below, only model B has nonzero $v_{aa}^{\rm sj}$ because its couplings involve spatial derivatives. The second contribution $v_{ba}^{\rm (sj)}$ takes the form (Eq.(S130))
\begin{equation}
   v_{ba,y}^{\rm (sj)} =A_y^{ba} (-i)\left(\tilde{\Pi}_-^{ab}(\ce_a)-\tilde{\Pi}_+^{ab}(\ce_a)\right)\,.
\end{equation}
Here $\tilde{\Pi}_\pm^{ab}(z)$ is the retarded/advanced phonon self-energy at complex frequency $z$ in the band diagonal basis, and $A_\mu^{ba}$ is the multi-band Berry connection of phonons (Eq.~(S113)).
This second contribution is similar in spirit to its  counterpart in the side-jump story of electrons \cite{Sinitsyn06}, in that it can be interpreted as a product of a scattering rate (the self-energies) and a coordinate shift (the multi-band Berry connection), although our expression for the coordinate shift is different from the electronic case.

Finally, the validity of the side-jump expression \eqref{eq:kappaH_semiclassical} is the same as the semi-classical Boltzmann equation, and we assume the lifetime $\Gamma_{\rm ph}$ is smaller than the temperature: \eqref{eq:kappaH_semiclassical} is obtained by performing the loop frequency integral in the Kubo formula expressions, and integral is controlled by the delta functions that correspond to the quasi-particle. Apart from that, \eqref{eq:kappaH_semiclassical} is general and can apply to different models. In particular, the phonon-defect resonances that we are going to discuss appear as delta functions in the self energies $\tilde{\Pi}^{ab}_\pm (z)$, and also in $v_{aa}^{\rm sj}$.

%Further details of \eqref{eq:kappaH_semiclassical} is discussed in the supplement \cite{Supp} (Eq. (S123)).

    Below we discuss the application of \eqref{eq:kappaH_semiclassical} to the three models A,B,C.

%%%%%%%%%%%%%%%%%%%%%%%%%%%%%%%%%%%%%%%
\subsection{Model A}

We consider a spin-1/2 defect in  an antiferromagnetic environment. As the defect is only sensitive to its nearby spins on the time scale of the defect dynamics, global magnetic order is not required. The defect spin is polarized by the local field as  %magnetic order as
\begin{equation}\label{eq:Afield}
  H_{\rm def}=-\frac{\Delta}{2}\sigma^3\,.
\end{equation}
We define the `3' axis as the direction of the {\it local\/} field on the impurity spin, with $\sigma^{1,2,3}$ obeying the algebra of Pauli matrices. The local field, $\Delta$, on the defect spin site is a combination of the effective field from the local magnetic order as well as the external magnetic field. Note that the orientation of the {\it applied\/} field will, in general, be different  from `3' axis because of the antiferromagnetic couplings, {\em e.g.\/} a N\'eel state becomes a canted state in the presence of an applied field, with the N\'eel order oriented orthogonal to the applied field, which changes the local field on an impurity spin. We always choose the applied field to be in the `$z$' direction (this could correspond to any crystallographic direction), and spatial co-ordinates so that the thermal Hall effect is measured in the $x$-$y$ plane, $\kappa_H \equiv \kappa_{xy}$. The spin-orbit interaction
leads to a spin-phonon coupling
\begin{equation}\label{eq:H_modelA}
  H_{\rm ph-def}=K_{i\alpha}\pi^i_o \sigma^\alpha\,,
\end{equation}
where $i=x,y,z$, $\alpha = 1,2,3$ is used for the spin components, the spin is at site $o$, and we are using different co-ordinate axes for the two indices of $K_{i\alpha}$. The Hamiltonian \eqref{eq:H_modelA} is a linear coupling between momentum and spin, similar to the purely electronic Rashba term \cite{Duine15}, and the couplings $K_{i\alpha}$ are similarly constrained by mirror plane symmetries near the impurity.
We find
\begin{eqnarray}
  \kappa_H &=&  \frac{m}{6 \pi N_{\rm sys}}\, \frac{  \Delta^4 }{  \Gamma_{ph} T^2 \sinh(\Delta/T)} \nonumber \\
  &~&~~~~~\times \left(\frac{1}{c_L} + \frac{1}{c_T} \right) \left(K_{x1} K_{y2} - K_{x2} K_{y1} \right)\,.
\label{eq:kappaH_modelA}
\end{eqnarray}
Here $N_{\rm sys}$ is the number of unit cells in the system, and for multiple non-interacting defects we multiply by the number of defects $N_i$. We can see that only couplings transverse to the defect polarization contributes to the thermal Hall effect (no coupling in \eqref{eq:kappaH_modelA} has the `3' index).

The last factor of \eqref{eq:kappaH_modelA} can be understood from symmetry considerations: The thermal Hall conductivity $\kappa_{xy}$ is invariant under spatial $\mathrm{SO}(2)_z$ along the z axis, and it is odd under spatial reflections $R_x, R_y$. These two conditions require a quadratic combination of $K_{i\alpha}$ with exactly one $x$ and one $y$ index, and is invariant under $\mathrm{SO}(2)_z$, yielding $K_{x\alpha}K_{x \beta}+K_{y \alpha} K_{y\beta}$ or $K_{x \alpha}K_{y\beta}-K_{y\alpha}K_{x\beta}$. Next, notice that the choice of the spin frame has a redundancy because we only fix the `$3$' direction to align with the polarization of the defect, and then the remaining `1' and `2' axes can rotate freely around the `3' axis. Physical observables should be independent of this free rotating angle, and only the coupling combination $K_{x1}K_{y2}-K_{x2}K_{y1}$ satisfies the requirement.

A sketch of the temperature dependence of \eqref{eq:kappaH_modelA} (with $\Gamma_{\rm ph}$ held at constant) is in Fig.~\ref{fig:kappa_sketch}, with a peak at $T \sim \Delta$.

%%%%%%%%%%%%%%%%%%%%%%%%%%%%%%%%%
\subsection{Model B}

The defect spin sitting at site $o$ couples to nearby lattice spins via exchange coupling
    \begin{equation}\label{eq:Hspin}
      H_{\rm def}^{\text{micro}}=\sum_{p}J_{po} S_{p}^{\alpha} S_{o}^{\alpha}\,,
    \end{equation}
and the summation is over lattice neighbors of $o$. As discussed in the supplement \cite{Supp} (Sec. 9), the magnetic order in the environment, and the dependence of the $J_{po}$ on the phonon displacement, leads to a $H_{\rm def}$ with an energy splitting $\Delta$ as in \eqref{eq:Afield}, and a defect-phonon coupling of the form
\begin{equation}
      H_{\rm ph-def}=K_{ij\alpha}\partial_i u^j_o \sigma^\alpha\,. \label{eq:kijk}
\end{equation}
The resulting thermal Hall effect is
\begin{equation}\label{eq:kappaH_modelB}
      \kappa_H =\frac{1}{30 \pi m N_{\rm sys}}\, \frac{  \Delta^4 }{  \Gamma_{ph} T^2 \sinh(\Delta/T)} \, \left(c_L^{-3}K_L+c_T^{-3}K_T\right)\,,
\end{equation}
where the coefficients $K_L$ and $K_T$ are
\begin{equation}\label{eq:K_L}
\footnotesize
    \begin{split}
K_L=&+2 \left(K_{xy2}+K_{yx2}\right) \left(K_{xx1}-K_{yy1}\right)-2 \left(K_{xy1}+K_{yx1}\right) \left(K_{xx2}-K_{yy2}\right)\\
&+K_{zx1} K_{zy2}-K_{zx2} K_{zy1}\\
&+K_{xz1} K_{yz2}-K_{xz2} K_{yz1}\\
&+K_{zx1} K_{yz2}-K_{zx2} K_{yz1}-K_{xz2} K_{zy1}+K_{xz1} K_{zy2}\,,
    \end{split}
    \end{equation}
\begin{equation}\label{eq:K_T}
\footnotesize
  \begin{split}
   K_T= &-\frac{5}{2} \left(\left(K_{xy1}+K_{yx1}\right) \left(K_{xx2}-K_{yy2}\right)-\left(K_{xy2}+K_{yx2}\right) \left(K_{xx1}-K_{yy1}\right)\right)\\
&+\frac{1}{2} \left(\left(K_{xy1}-K_{yx1}\right) \left(K_{xx2}+K_{yy2}\right)-\left(K_{xy2}-K_{yx2}\right) \left(K_{xx1}+K_{yy1}\right)\right)\\
&+K_{zx1} K_{zy2}-K_{zx2} K_{zy1}\\
&+4 K_{xz1} K_{yz2}-4 K_{xz2} K_{yz1}\\
&+K_{zz1} \left(K_{xy2}-K_{yx2}\right)+K_{zz2} \left(K_{yx1}-K_{xy1}\right)\,.\\
  \end{split}
\end{equation}
Similar to model A,  each line of $K_L$ and $K_T$ are quadratic combinations of couplings which has the same transformation property as $\kappa_{xy}$ and is invariant under rotation around the `3' direction in spin indices.
%where $K_L$ and $K_T$ are certain quadratic combinations of the coupling constants $K_{ij\alpha}$, which can be partially understood from symmetry perspective \cite{Supp}.
The form of \eqref{eq:kappaH_modelB} is such that a nonzero thermal-Hall effect requires a non-trivial non-coplanar spin order near the impurity site. In particular, a conventional canted N\'eel order does not lead to a non-zero thermal Hall conductivity. A thermal Hall proportional to external field can be derived from a combination of coplanar non-collinear magnetic order and additional canting due to external field \cite{Supp}.
Since a finite thermal-Hall effect in model B does not require spin-orbit coupling, it makes it an attractive candidate for the thermal-Hall observed in pseudogap cuprates. As mentioned earlier, the requirement of a non-trivial non-coplanar local spin order seems to be satisfied in underdoped cuprates \cite{Julien19}.

In passing, we also mention that depending on the system the coupling of defect spin to its neighbors may not be just Heisenberg type. In Eq. (\ref{eq:Hspin}) one could also consider additional explicit SU$(2)$ symmetry breaking terms such as Dzyloshinskii-Moriya interaction resulting from spin-orbit coupling. In this case, a non-trivial non-coplanar local spin order is not required, and a conventional canted N\'eel order may also lead to finite thermal Hall.
% The form of Eq.~\eqref{eq:kappaH_modelB} is such that a nonzero thermal Hall effect will require either non-coplanar spin order near the impurity, or spin-orbit coupling. A thermal Hall proportional to external field can be derived from a combination of coplanar magnetic order and additional canting due to external field \cite{Supp}.

Finally, when couplings from both model A and model B appear at the same time, we found that the thermal Hall responses simply add up without interference terms.

% \begin{comment}
% \begin{widetext}
%     \begin{equation}\label{eq:K_T}
%       \begin{split}
%           K_T =&-\frac{5}{2}\left[\left(K_{{xx1}}-K_{{yy1}}\right) \left(K_{{xy2}}+K_{{yx2}}\right)-\left(K_{{xx2}}-K_{{yy2}}\right) \left(K_{{xy1}}+K_{{yx1}}\right)\right]\\
%            &+\frac{1}{2}\left[\left(K_{{xx1}}+K_{{yy1}}\right) \left(K_{{xy2}}-K_{{yx2}}\right)-\left(K_{{xx2}}+K_{{yy2}}\right) \left(K_{{xy1}}-K_{{yx1}}\right)\right]\\
%           &+K_{{zz1}} \left(K_{{yx2}}-K_{{xy2}}\right)
%           +K_{{zz2}} \left(K_{{xy1}}-K_{{yx1}}\right)\\
%           &-4 K_{{xz1}} K_{{yz2}}+4 K_{{xz2}} K_{{yz1}}\\
%           &-K_{{zx1}} K_{{zy2}}+K_{{zx2}} K_{{zy1}}\,,
%       \end{split}
%     \end{equation}
%     \begin{equation}\label{eq:K_L}
%       \begin{split}
%          K_L  =& -2 \left(K_{xx1}-K_{yy1}\right) \left(K_{xy2}+K_{yx2}\right)+2 \left(K_{xx2}-K_{yy2}\right) \left(K_{xy1}+K_{yx1}\right)\\
%          &-\left(K_{xz1}+K_{zx1}\right) \left(K_{yz2}+K_{zy2}\right)+\left(K_{xz2}+K_{zx2}\right) \left(K_{yz1}+K_{zy1}\right)\,.
%       \end{split}
%     \end{equation}
% \end{widetext}
% \end{comment}

%%%%%%%%%%%%%%%%%%%%%%%%%%%%%%%%%%%%%%%%%%
\subsection{Model C}

This model concerns an orbital defect consisting of a singlet and a triplet:
      \begin{equation}\label{}
      H_{\rm def}=\sum_{\ell=0}^{1}\sum_{m=-\ell}^{\ell} E_{\ell m}\ket{ \ell m}\bra{ \ell m}\,,
    \end{equation}
The ground state has energy $E_{00}=0$, and the excited triplet is split by Zeeman field $E_{1m}=\Delta-m\Delta_Z$, $\Delta\gg \Delta_Z$. We assume the phonon-defect coupling to be
\begin{equation}\label{eq:ph-def_coupling}
      H_{\rm ph-def}= \gamma \, {\pi}_o^i \,  \delta{V}^i\,.
\end{equation}
The defect is located at site $o$ and $V^i$ is a vector operator of the defect system, and $\delta V^i$ is the deviation from equilibrium. We have assumed the operator $V$ describes orbital effects, and hence shares the same index as $u$ and $\pi$.
 %We have assumed that the thermal equilibrium value of $\braket{V^i}$ only shifts the ion equilibrium positions, and by switching to the shifted coordinate system we replace $V^i$ by $\delta V^i=V^i-\braket{V^i}$.
 The defect can be modelled as trapped particle with momentum $V^i$, and a coupling of the form $V^i {\pi}_o^i$ is generated by a canonical transformation described in Ref.~\cite{sun2021}. The matrix elements of the vector operator $V^i$ are described by Wigner-Eckart theorem: let $V^0=V^z$, $V^{\pm}=\mp(V^x\pm i V^y)/\sqrt{2}$, the matrix element is given by
\begin{equation}\label{}
      \braket{\ell m|V^q|\ell'm'}=\Braket{\ell||V||\ell'}\Braket{\ell'm'1q|\ell m}\,;
\end{equation}
on the RHS, the first term is the reduced matrix element and the second term is the Clebsch-Gordan coefficient.

To linear order in the Zeeman splitting $\Delta_Z$, we found:
    %\begin{equation}\label{eq:kappaH_model1}
   %   \kappa_{H}=-\frac{a^3}{L^3}\frac{\beta^2 \gamma ^2 \Delta ^4  m \left(c_L^{-1}+c_T^{-1}\right) }{6 \pi  \Gamma _{ph}\sinh(\beta \Delta)}\,,
  %  \end{equation}
\begin{equation}\label{eq:kappaH_modelC}
      \kappa_{H}=\frac{1}{N_{\rm sys}}\frac{\alpha _R \gamma ^2 \Delta_Z  m \Delta^3 \left(e^{\Delta/T} \left(  \Delta-4T\right)+4T\right) (c_L^{-1}+c_T^{-1})}{12 \pi  \text{$\Gamma_{\rm ph} $} \left(e^{\Delta/T}+3\right) T^3 \sinh^2(\Delta/2T)}\,.
\end{equation}
Here  $\alpha_R$ is related to the reduced matrix elements of the model, given by the expression
\begin{equation}\label{}
       \alpha_R=\Braket{1||V||0}\Braket{0||V||1}/\sqrt{3}\,.
\end{equation} We can see that \eqref{eq:kappaH_modelC} arises from phonons resonating with transitions between the two multiplets. The sign change of \eqref{eq:kappaH_modelC} at low temperature is an artifact of our defect model and is not universal. There are also contributions due to resonance with transitions within the excited triplet, which is similar to the two-level models considered before, but suppressed by a thermal weight $e^{- \Delta/T}$ and higher power of $\Delta_Z/\Delta$.
\begin{figure}
  \centering
  \subfloat[Models A and B]{\includegraphics[width=0.25\textwidth]{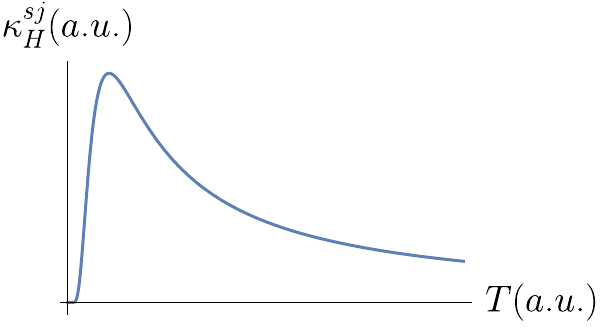}} ~~~
  \subfloat[Model C]{\includegraphics[width=0.25\textwidth]{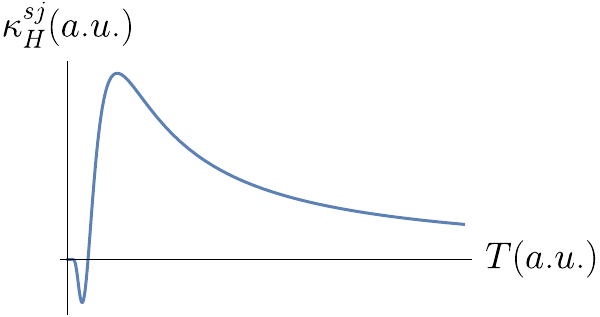}}
  % \begin{subfigure}[t]{0.25\columnwidth}
  % \includegraphics[width=\textwidth]{Fig3a.pdf}
  % \caption{Models A and B}
  % \end{subfigure}
  % \begin{subfigure}[t]{0.25\columnwidth}
  % \includegraphics[width=\textwidth]{Fig3b.pdf}
  % \caption{Model C}
  % \end{subfigure}
  \caption{Temperature dependence of $\kappa_H$ for a constant $\Gamma_{ph}$ for (a) models A and B, and (b) model C. Note that $\kappa_{H}$ for models A and B have similar temperature dependence. However, all the models correspond to {\em different} physical situations as discussed in the text. But the underlying mechanism of phonon thermal-Hall effect is the same. The $\kappa_{H}$ for different models is {\em not} to be compared with each other.  }\label{fig:kappa_sketch}
\end{figure}

%%%%%%%%%%%%%%%%%%%%%%%%%%%%%%%%%%%%%%%%%%%%%%%%%%%%%%%%%%
\section{Connections to observations}

We now comment on various connections between our results and observations on different materials. We will be focusing on model B which is most relevant to cuprate experiments.

\subsection{Connection to spin glass order}

    It is shown \cite{Supp} that a sufficient condition for the thermal Hall effect
    in the absence of spin-orbit coupling is a non-coplanar magnetic order around the defect spin. We relate this to the observation of spin glass order in pseudogap cuprates \cite{Julien19}, which emerges exactly at the critical doping. Although our model B is formulated in the context of a single defect in a magnet, it is not sensitive to the global magnetic order and therefore it can also be interpreted as a model characterizing phonon-spin scattering in a random environment, including spin glass. From this perspective, the emergence of phonon thermal Hall effect in the pseudogap is a consequence of the non-coplanar magnetism due to spin glass. The spin glass picture also explains the similarity between $\kappa_{xy}$ and $\kappa_{yz}$: In a random ensemble of spins without spin-orbit coupling, there is no difference between the $z$ direction or the $x$ direction.

\subsection{Hall angle}
As we discussed in the introduction, a feature seen in the cuprates \cite{Grissonnanche2019,Grissionnanche2020,Boulanger2020,boulanger2021} and magnetic insulators \cite{chen2021large}
is that the Hall angle $|\kappa_H/\kappa_{xx}|$ remains at the order $10^{-3}$ at temperature $\sim 20\rm{K}$ and magnetic field $\sim 15 \rm{T}$ across different materials despite drastic variation of $\kappa_{xx}$ (for a nice summary, see Table.I of Ref.~\cite{boulanger2021}). In our theory, this ratio is naturally independent of $\Gamma_{\rm ph}$. In particular, for resonant scattering, our theory predicts that this ratio contains a peak at a $T$ of order $\Delta$, and this peak should have different physical origin from the peak in $\kappa_{xx}$. Let us contrast to the case where skew scattering dominates $\kappa_H$. As is pointed out in Ref.~\cite{Balents2022}, in this case $\kappa_H\propto 1/\Gamma_{ph}^2$ and therefore the Hall angle in general can depend on $\Gamma_{\rm ph}$. Instead, for skew scattering the Hall resistivity $\kappa_{H}/\kappa_{xx}^2$ is needed to strip off effects of the phonon mean-free path. The only exception is when the skew-scattering channel also dominates over other ordinary scattering \cite{kivelson20}, then $|\kappa_H/\kappa_{xx}|$ would be independent of phonon mean-free path, but this is not likely the case in cuprates.

\subsection{Numerical estimates for cuprates}
\label{ThetaH}
%A notable feature of the recent observations of $\kappa_H$ in the cuprates \cite{Grissionnanche2020,Boulanger2020,boulanger2021} is that the enhanced phonon contribution appears only in the pseudogap regime, where quasi-static magnetic moments are likely to be present \cite{Julien19}. Our model B is an attractive description of phonon-defect dynamics in such a regime, given its isotropic Zeeman coupling to the applied field, and its non-zero thermal Hall response even in the absence of spin-orbit coupling.
When there is no external magnetic field, we assume the total thermal conductivity, after averaging over a large number of spins, vanishes due to the fact that each spin polarizes in a random direction. With an external magnetic field $B$ applied, the spins will cant according to $B$. Since our convention is to fix the polarization of spins to be along the `3' direction, the leading order effect of $B$ is to modify the coupling tensor $K_{ij\alpha}$ with a term proportional to the canting angle $\theta_c\sim \Delta_B/|K_{ij\alpha}|$, where $\Delta_B=2\mu_B B$ is the Zeeman energy. This additional coupling will give rise to a thermal Hall effect which is linear in $B$.

In Section 10 of the supplement \cite{Supp}, we performed an order-of-magnitude estimate of the Hall angle, with the result
\begin{eqnarray}\label{eq:ratio_expression}
    |\Theta_H| &=&   \frac{|\kappa_H|}{\kappa_{xx}}=\frac{5 \Delta _B k_B^3 T_D^3}{4 \pi ^4 c^5 \rho  \hbar ^3}A_H \frac{N_i}{N_{sys}}\Phi\left(\frac{\Delta}{k_B T}\right)\nonumber \\
    \Phi(x)& =& \frac{x^5}{\sinh(x)}\,.
\end{eqnarray}
Here, we have used the fact that microscopically both the defect spin splitting and the defect-phonon coupling originate from spin exchange $J$, and can be represented by $\Delta$, with a numerical coefficient absorbed in $A_H$. $A_H$ is a numerical coefficient depending on the tensor structure of the couplings and the ratio $\Delta/J$. The factor $N_i/N_{sys}$ is the concentration of spin defects participating in phonon scattering. For cuprates, typically sound velocity $c=5000\rm{m/s}$, density $\rho=6000 \rm{kg/m^3}$, Debye temperature $T_D=400\rm{K}$. It is a remarkable fact that $\Delta$ and $1/T$ appear with the same power in the prefactor of (\ref{eq:ratio_expression}), and so
the dependence on $\Delta$ and $T$ can be encapsulated into a scaling function $\Phi$, implying the maximal Hall angle is \emph{independent} of the magnitude of microscopic exchange coupling. The maximal Hall angle is achieved at $T\approx 0.2 \Delta$, with the value
\begin{equation}\label{eq:ratio_result}
      \left(\frac{\kappa_H(B=15\rm{T})}{\kappa_{xx}}\right)_{max}=1.2\times 10^{-3}A_H \frac{N_i}{N_{sys}}\,.
\end{equation} For phonons scattering off glassy quasi-static spin order, we assume $N_i/N_{sys}$ to be order one. In experiments the peak is around $20\rm{K}$, from which we can determine $\Delta\approx 100\rm{K}$. Note that this value is at the same order as the Neel temperature of cuprates, which is around 300-500K. This is expected because they share the same microscopic origin, but $\Delta$ can be smaller from non-collinearity of neighboring spins. It is also notable in electron-doped cuprates the Hall angle increases with doping \cite{boulanger2021}.

\subsection{Phonon lifetime}
We have taken the phonon lifetime $\tau_{\rm ph} = 1/\Gamma_{\rm ph}$ as an fixed parameter, but $\Gamma_{\rm ph}$ could have additional $T$ and $B$ dependence. In Fig.~\ref{fig:kappa_sketch}, $\kappa_H$ is plotted with $\Gamma_{\rm ph}$ fixed. In the cuprates, $\Gamma_{\rm ph}$ arises from other scattering mechanisms \cite{singh2020}, and the resonant scattering we have considered
for $\kappa_H$ is not likely to be the dominant mechanism for $\Gamma_{\rm ph}$: resonant scattering suppresses the longitudinal conductivity $\kappa_{xx}$, but experimentally $\kappa_{xx}$ and $\kappa_{H}$ are found to peak at about the same temperature \cite{Grissionnanche2020,Boulanger2020,chen2021large,boulanger2021}.

\subsection{Skew scattering}
Skew scattering is a subdominant contribution in our model. Diagrammatically, this comes from contracting the 4 $\sigma^\alpha$ correlation function with phonon legs; keeping only the intraband component of energy current vertex, this contribution becomes proportional to $K^4/\Gamma_{\rm ph}^2$.  As we have discussed, in the cuprates $\Gamma_{\rm ph}$ arises from other mechanisms (i.e. independent of $K$), and so this contribution is of order $K^4$, higher order than that in (\ref{eq:kappaH_modelB}). Furthermore, the phonon-defect coupling (\eqref{eq:kijk}) only involves phonon wave functions of the same parity, and the skew scattering contribution is expected to vanish following the same argument as \cite{Guo2021}.

\subsection{Average over defects}
Our calculation  picked up the resonant contribution to thermal Hall conductivity due to a single defect, which is exponentially suppressed at low temperature $T\ll \Delta$. A power-law decay at low $T$ can be obtained by an average over defects with a distribution of values of $\Delta$ and $K_{ij\alpha}$, if there exists a power law tail in the distribution of $\Delta$.

At low $T$, this resonant side-jump contribution might also be smaller than other mechanisms such as Berry curvature, which shows a power law in $T$ \cite{kivelson20,ZhangTeng21,ye2021phonon}.

\subsection{Comparison to charge Hall effect}

    It is also interesting to compare our result to charge Hall effect in electronic systems, e.g. graphene \cite{Sinova07}. The major difference is the source of chirality for quasiparticles. In graphene, the electrons gain chirality from the intrinsic Berry curvature, and even scattering with ordinary disorders will inherit the chirality. Therefore, all scattering mechanisms for electron contribute to both the total lifetime $\tau_{el}$ and the skew lifetime $\tau_{skew}$, and this makes skew scattering the dominant mechanism.
    In contrast, we started from non-chiral acoustic phonons which scatter both on non-chiral agents (represented by $\Gamma_{\rm ph}$) and spin defects, which makes side jump the dominant contribution.

\subsection{Spin ice}
In a recent experiment on metallic spin ice $\rm{Pr}_2\rm{Ir}_2\rm{O}_7$ \cite{Machida2022}, a giant thermal Hall effect has also been attributed to phonons with a similar Hall angle. In this material, the low-energy non-Kramers doublet is a pseudo-spin 1/2 derived from crystal electric field, which could also yield coupling to phonons without time-reversal symmetry breaking. There, the phonon-pseudo spin coupling not only suppresses longitudinal heat conduction but also contributes a thermal Hall effect. Our model B is expected to apply when $B\parallel [001]$, where paramagnetic pseudo spin-phonon scattering is present. The difference from cuprates is that the phonon decay rate $\Gamma_{\rm ph}$ is now dominated by phonon-pseudo spin  scattering. Application of our theory to $\rm{Pr}_2\rm{Ir}_2\rm{O}_7$ will be discussed in a subsequent paper.

\acknow{We thank G.~Grissonnanche, S.~Kivelson, Jing-Yuan Chen, L.~Spodyneiko, and L.~Taillefer for valuable discussions. This research was supported by National Science Foundation grant No. DMR-2002850, by the Simons Collaboration on Ultra-Quantum Matter which is a grant from the Simons Foundation (651440, S.S.).
D.G.J acknowledges support from the Leopoldina fellowship by the German National Academy of Sciences through grant no. LPDS 2020-01. H.G. was supported in part by the Heising-Simons Foundation, the Simons Foundation, and National Science Foundation Grant No. NSF PHY-1748958.}

\showacknow{} % Display the acknowledgments section

% Bibliography
\bibliography{phonon}



\section*{Supplementary material}

**Supporting Information Text**

**Contents**

In this supplement material, we present detailed calculations of the thermal Hall effects discussed in the main text. We will discuss model C in Part I because some of the expressions are simpler in this model due to the simplicity of phonon-defect coupling. We later generalize the formalism to account for the more complicated couplings in models A and B in Part II. We also include a calculation that reproduces the Berry curvature formula of phonon Hall effect for quadratic phonon systems in Part III.

# Part I

# Model C

In this part, we evaluate the thermal Hall effect of model C by keeping all contributions to the modified Kubo formula.

## 1. Quadratic Phonon Hamiltonian

In this part, we discuss properties of the quadratic phonon Hamiltonian.

In the main text, the quadratic phonon Hamiltonian without dissipation is given by

$$H_{ph} = \sum_p \frac{\pi_p^i \pi_p^i}{2m} + \frac{1}{2} \sum_{pq} u_p^i C_{pq}^{ij} u_q^j, \qquad [\text{S1}]$$

where $m$ is the ion mass and $C_{pq}^{ij}$ is the elastic matrix, and the variables $u_p^i$ and $\pi_p^i$ satisfy the commutation relation

$$[u_p^i, \pi_q^j] = i\delta^{ij}\delta_{pq}. \qquad [\text{S2}]$$

It is convenient to treat $u$ and $\pi$ on equal footing, by introducing the variable $\zeta_p^I = (u_p^i, \pi_p^i)$. Then any quadratic phonon Hamiltonian can be written as

$$H = \frac{1}{2} \sum_{pq} \zeta_p^I h^{IJ}(p,q) \zeta_q^J, \qquad [\text{S3}]$$

where $\zeta$'s satisfy the canonical commutation relation

$$[\zeta_p^I, \zeta_q^J] = iJ^{IJ}(p,q). \qquad [\text{S4}]$$

Here repeated $I, J, \ldots$ indices are Einstein summed. Assuming $\zeta$'s are Hermitian, then the matrix $h$ is Hermitian and $J$ is real antisymmetirc. Furthermore, by adding a suitable constant to $H$ we can make $h$ real and symmetric.

For the Hamiltonian Eq. (S1), we have explicitly

$$h^{IJ}(p,q) = \begin{pmatrix} C_{pq}^{ij} & 0 \\ 0 & \frac{1}{m}\delta^{ij}\delta_{pq} \end{pmatrix}, \qquad [\text{S5}]$$

and

$$J^{IJ}(p,q) = \begin{pmatrix} 0 & \delta^{ij}\delta_{pq} \\ -\delta^{ij}\delta_{pq} & 0 \end{pmatrix} \qquad [\text{S6}]$$

respectively.

2

**A. Diagonalization.** First we diagonalize the Hamiltonian Eq. (S3). For notational simplicity, in this subsection we group the indices together, writing $\zeta_p^I = \zeta_a$. The starting point is

$$H = \frac{1}{2} \sum_{ab} \zeta_a h_{ab} \zeta_b, \quad [\zeta_a, \zeta_b] = i J_{ab}. \tag{S7}$$

The Heisenberg equation reads

$$\frac{d\zeta_a}{dt} = i[H, \zeta_a] = (Jh\zeta)_a. \tag{S8}$$

The normal modes satisfy the eigenvalue equation

$$\omega_a \psi_a = iJh\psi_a. \tag{S9}$$

Suppose the matrix $iJh$ is diagonalized by matrix $M$:

$$iJhM = M\mathcal{E}, \tag{S10}$$

where $\mathcal{E}$ is a diagonal matrix. Multiply $(iJ)^{-1}$ on both sides of Eq. (S10), and contract with $M^\dagger$, we obtain

$$M^\dagger h M = M^\dagger (iJ)^{-1} M \mathcal{E}. \tag{S11}$$

Taking the diagonal elements of the above equation, and using the fact that $h, iJ$ are Hermitian, we conclude that $\mathcal{E}$ is real. Because $h$ is positive definite, $\mathcal{E}$ is nonzero.

Taking the Hermitian conjugation of Eq. (S11), we obtain

$$M^\dagger (iJ)^{-1} M \mathcal{E} = M^\dagger h M = \mathcal{E} M^\dagger (iJ)^{-1} M. \tag{S12}$$

Here we have used the fact that $\mathcal{E}$ is real. Therefore $M^\dagger (iJ)^{-1} M$ can also be chosen to be diagonal. The positive definiteness of $h$ implies that $M^\dagger (iJ)^{-1} M$ has the same sign as $\mathcal{E}$, therefore we normalize it as

$$M^\dagger (iJ)^{-1} M = \operatorname{sgn} \mathcal{E}. \tag{S13}$$

Because $J, h$ are real, the complex conjugate of Eq. (S10) yields

$$iJhM^* = -M^*\mathcal{E}. \tag{S14}$$

This implies that eigenvalues of $\mathcal{E}$ are paired up in opposite signs.

Let's define a particle conjugation operator $\mathcal{C}$, which satisfies

$$M^* = M\mathcal{C}. \tag{S15}$$

The definition implies

$$\mathcal{C}^* = \mathcal{C}^{-1}. \tag{S16}$$

From Eq. (S14), we find

$$\mathcal{C}\mathcal{E} = -\mathcal{C}\mathcal{E}, \tag{S17}$$

i.e. $\mathcal{C}$ exchange modes of opposite energy.

Comparing Eq. (S13) with its complex conjugation, we can show that

$$\mathcal{C}^T = \mathcal{C}^{-1}. \tag{S18}$$

The new normal modes are given by

$$v = M^{-1}\zeta.  \quad [S19]$$

Their commutation relation is given by

$$K_{ab} \equiv [v_a, v_b], \quad K = M^{-1} iJ (M^T)^{-1}. \quad [S20]$$

By construction $K$ is an antisymmetric matrix.

Multiplying Eq. (S10) with $(iJ)^{-1}$, and contract with $M^T$, we obtain

$$K^{-1}\mathcal{E} = M^T h M = -\mathcal{E} K^{-1}. \quad [S21]$$

Therefore $K$ anticommutes with $\mathcal{E}$. Furthermore, we can relate $K$ to $\mathcal{C}$ via the normalization condition Eq. (S13):

$$K = \operatorname{sgn} \mathcal{E} \mathcal{C}^T. \quad [S22]$$

Since $K$ is antisymmetric, we can derive that

$$\mathcal{C} = \mathcal{C}^T = \mathcal{C}^* = \mathcal{C}^{-1}. \quad [S23]$$

Consider the subspace spanned by the two eigenvectors of opposite eigenvalue, in this space $\operatorname{sgn}\mathcal{E} = \sigma^z$ and therefore $\mathcal{C} = \sigma^x$, $K = i\sigma^y$. Therefore, the normal mode corresponding to positive energy can always be defined as annihilation operator, while the normal mode corresponding to negative energy can be defined as creation operator, and they satisfy the commutation relation.

In terms of the new normal modes, the Hamiltonian reads

$$\tilde{h} = M^T h M = \mathcal{E} K. \quad [S24]$$

Expanding in the subspace of conjugate normal modes, the Hamiltonian is equivalent to the standard boson Hamiltonian $(|\omega|/2)(aa^\dagger + a^\dagger a)$.

**B. Free Phonon Green's Function.** The phonon Green's function in imaginary time is defined as

$$D_{ab}(\tau) = -T_\tau \langle \zeta_a(\tau) \zeta_b(0) \rangle. \quad [S25]$$

This Green's function can be calculated by using the diagonalization method derived above. Expand $\zeta$ in terms of the creation and annihilation operators, evaluate the expectations and using Eq. (S20), we obtain

$$D(i\omega_n) = \frac{1}{i\omega_n - iJh} iJ. \quad [S26]$$

The above result can be alternatively derived from phase space path integral. For a 1D particle with momentum $p$ and coordinate $q$, its imaginary-time path integral is given by

$$\int Dp Dq \exp\left(-\int d\tau \left(-ip\frac{dq}{d\tau} + H(p,q)\right)\right). \quad [S27]$$

Generalizing the above form many variables, we just need to find the canonical momentum $P_a$ for $\zeta_a$, which satisfies $[P_a, \zeta_b] = -i\delta_{ab}$. This condition is solved by $P_a = -(J^{-1}\zeta)_a$. The path integral for the phonon system is therefore

$$\int D\zeta \exp\left(-\int d\tau \left(\frac{1}{2}\zeta^T (iJ)^{-1} \frac{\partial \zeta}{\partial \tau} + \frac{1}{2}\zeta^T h \zeta\right)\right), \quad [S28]$$

where we have inserted a factor $1/2$ to avoid double counting. Switching to frequency space, this immediately yields the Green's function Eq. (S26).

In actual computations of the thermal Hall conductivity, we need the retarded and advanced Green's functions with dissipation terms put in by hand:

$$D_\pm(z) = \frac{1}{z \pm \frac{i\Gamma_{\text{ph}}}{2} - iJh} iJ. \quad [S29]$$

**C. Acoustic Phonon.** For our problem, we assume the Hamiltonian describes acoustic phonons in an isotropic lattice. The elastic matrix $C_{pq}^{ij}$ can be fourier transformed to momentum space by $C^{ij}(k) = (1/N_{cell}) \sum_p C_{pq}^{ij} e^{-ik \cdot (p-q)}$ where $N_{cell}$ is the number of unit cells. In momentum representation, we have

$$C^{ij}(k) = mc_T^2 \delta^{ij} k^2 + m(c_L^2 - c_T^2) k^i k^j, \quad [S30]$$

where $c_L$ is the longitudinal velocity and $c_T$ is the transverse velocity. The momentum space representation of $h$ and $J$ are

$$h^{IJ}(k) = \begin{pmatrix} C^{ij}(k) & 0 \\ 0 & 1 \end{pmatrix}, \quad [S31]$$

$$J^{IJ}(k) = \begin{pmatrix} 0 & \delta^{ij} \\ -\delta^{ij} & 0 \end{pmatrix}. \quad [S32]$$

The Hamiltonian can now be diagonalized for each individual $k$. An explicit solution for the matrix $M(k)$ is

$$M(k) = \begin{pmatrix} \begin{pmatrix} \frac{e_k^1}{\sqrt{m\omega_k^1}} & \frac{e_k^2}{\sqrt{m\omega_k^2}} & \frac{e_k^3}{\sqrt{m\omega_k^3}} \end{pmatrix} & 0 \\ 0 & \begin{pmatrix} e_k^1 \sqrt{m\omega_k^1} & e_k^2 \sqrt{m\omega_k^2} & e_k^3 \sqrt{m\omega_k^3} \end{pmatrix} \end{pmatrix}$$
$$\times \left( \frac{1}{\sqrt{2}} \begin{pmatrix} 1 & 1 \\ -i & i \end{pmatrix} \otimes I_3 \right). \quad [S33]$$

The eigenvalues are

$$\mathcal{E} = \mathrm{diag}(\omega_k^1, \omega_k^2, \omega_k^3, -\omega_k^1, -\omega_k^2, -\omega_k^3). \quad [S34]$$

Here $e_k^a$ is a 3 by 1 column vector that describes the polarization vector of the $a$-th mode, and $\omega_k^a$ is the corresponding frequency. $I_3$ is the 3 by 3 identity matrix. From Eq. Eq. (S30) we have one longitudinal mode $\omega_k^3 = c_L k$ and two transverse modes $\omega_k^{1,2} = c_T k$.

## 2. Defect Green's function

In this part we compute the defect Green's function which will be used to compute the thermal Hall effect. We assumed the defect-phonon coupling to be of the form

$$H_{ph-def} = \gamma \vec{\pi}_o \cdot \delta \vec{V}. \quad [S35]$$

Here we have subtracted off the equilibrium value $\delta \vec{V} = \vec{V} - \langle \vec{V} \rangle$ to fix the equilibrium position of $\vec{\pi}$ at zero. Since the defect is localized at the site $o$ and couples only to momentum, the defect Green's function, when written in terms of the index structure of $\zeta_p^I$, is given by

$$S_{pq}^{IJ}(i\omega_n) = \delta_{po} \delta_{qo} \begin{pmatrix} 0 & 0 \\ 0 & S^{ij}(i\omega_n) \end{pmatrix}, \quad [S36]$$

where $S^{ij}(i\omega_n)$ is the Green's function written in the defect Hilbert space

$$S^{ij}(i\omega_n) = -\int_0^\beta d\tau e^{i\omega_n \tau} T_\tau \langle V^i(\tau) V^j(0) \rangle. \quad [S37]$$

Here in the definition we have used $V^i$ instead of the subtracted version $\delta V^i = V^i - \langle V^i \rangle$, because ultimately we only need the retarded and the advanced Green's functions, which are insensitive to variable shifts.

In the continuum limit, it's convenient to switch to momentum space, then the defect Green's function becomes

$$\langle k_1|S^{IJ}(i\omega_n)|k_2\rangle = \frac{1}{N_{\text{sys}}}\begin{pmatrix} 0 & 0 \\ 0 & S^{ij}(i\omega_n) \end{pmatrix}. \quad [S38]$$

Here we have assumed the momentum eigenstates are normalized to one $\langle k_1|k_2\rangle = \delta_{k_1,k_2}$, and therefore the normalization constant is given by the number of unit cells $N_{\text{sys}}$.

We do not include dissipation $\Gamma_s$ for the defect Green's function. Because we found the thermal Hall conductivity is smooth in the $\Gamma_s \to 0$ limit, and therefore the effect of $\Gamma_s$ will be higher order corrections in terms of the coupling $\gamma$.

Now we compute $S^{ij}$. The model is a 4-level system with a Zeeman field:

$$H_{def} = \sum_{l=0}^{1} \sum_{m=-l}^{l} E_{lm} |lm\rangle \langle lm|, \quad [S39]$$

The ground state has energy $E_{00} = 0$, and the excited triplet is split by Zeeman field $E_{1m} = \Delta - m\Delta_Z$, $\Delta \gg \Delta_Z$.

The retarded Green's function is given by

$$S_+^{ij}(t) = -i\theta(t)\langle[V^i(t), V^j(0)]\rangle. \quad [S40]$$

To compute the Green's function, it is convenient to switch to the spherical basis

$$\begin{pmatrix} V^x \\ V^y \\ V^z \end{pmatrix} = \underbrace{\begin{pmatrix} \frac{-1}{\sqrt{2}} & \frac{1}{\sqrt{2}} & 0 \\ \frac{i}{\sqrt{2}} & \frac{i}{\sqrt{2}} & 0 \\ 0 & 0 & 1 \end{pmatrix}}_{\Lambda^{iq}} \begin{pmatrix} V^+ \\ V^- \\ V^0 \end{pmatrix}. \quad [S41]$$

The correlator for the spherical operators can be computed as $(q, q' = +, -, 0)$:

$$\langle V^q(t)V^{q'}(0)\rangle = \sum_{lm,l'm'} \frac{e^{-\beta E_{lm}}}{Z} e^{-i(E_{l'm'}-E_{lm})t} \langle lm|V^q|l'm'\rangle \langle l'm'|V^{q'}|lm\rangle$$

$$= \sum_{lm,l'm'} \frac{e^{-\beta E_{lm}}}{Z} e^{-i(E_{l'm'}-E_{lm})t} \langle l\|V\|l'\rangle \langle l'\|V\|l\rangle \langle l'm'1q|lm\rangle \langle lm1q|l'm'\rangle. \quad [S42]$$

In the second line, we have used the Wigner-Eckart theorem to factorize the matrix elements into Clebsch-Gordon coefficients and the reduced matrix elements $\langle l'\|V\|l\rangle$.

For the 4-level model, we find

$$S_\pm^{ij}(z) = \alpha_R K_1^{ij}(z_\pm) + \alpha_R K_2^{ij}(z_\pm) + \beta_R K_3^{ij}(z_\pm). \quad [S43]$$

The two constants are

$$\alpha_R = \frac{1}{\sqrt{3}}\langle 1\|V\|0\rangle\langle 0\|V\|1\rangle, \quad [S44]$$

$$\beta_R = \langle 1\|V\|1\rangle^2. \quad [S45]$$

The functions involved are

$$K_1^{ij}(z) = -\frac{e^{\beta(\Delta_Z+\Delta)} - 1}{1 + e^{\beta\Delta_Z} + e^{2\beta\Delta_Z} + e^{\beta(\Delta_Z+\Delta)}} F^{ij}(z, \Delta_Z + \Delta), \quad [S46]$$

$$K_2^{ij}(z) = -\frac{e^{2\beta\Delta_Z} - e^{\beta(\Delta_Z+\Delta)}}{1 + e^{\beta\Delta_Z} + e^{2\beta\Delta_Z} + e^{\beta(\Delta_Z+\Delta)}} F^{ij}(z, \Delta_Z - \Delta), \quad [S47]$$

6

$$K_3^{ij}(z) = -\frac{e^{2\beta\Delta_Z} - 1}{1 + e^{\beta\Delta_Z} + e^{2\beta\Delta_Z} + e^{\beta(\Delta_Z+\Delta)}} F^{ij}(z, \Delta_Z),  \qquad [S48]$$

$$F^{ij}(z, \epsilon) = \begin{pmatrix} \frac{\epsilon}{z^2-\epsilon^2} & \frac{iz}{z^2-\epsilon^2} & 0 \\ \frac{-iz}{z^2-\epsilon^2} & \frac{\epsilon}{z^2-\epsilon^2} & 0 \\ 0 & 0 & 0 \end{pmatrix}. \qquad [S49]$$

The contribution of $K_3$ to thermal Hall effect is very similar to the result of the two-level model discussed later. Compared to $K_1$ and $K_2$, $K_3$ is smaller by powers of $\Delta_Z/\Delta$ ($\Delta_Z \ll \Delta$) as well as a thermal weight $e^{-\beta\Delta}$, therefore we will only keep $K_1$ and $K_2$ in the later calculations.

## 3. Energy Current

In this part, we shall compute the energy current operator. The energy current receives contribution from the quadratic phonon Hamiltonian and the phonon-defect coupling.

**A. Quadratic Phonon Part.** From Eq. Eq. (S3), we can derive the Hamiltonian density to be

$$H_p = \frac{1}{4} \sum_m \left( \zeta_p^I h^{IJ}(p,m) \zeta_m^J + \zeta_m^J h^{JI}(m,p) \zeta_p^I \right). \qquad [S50]$$

For a local Hamiltonian $h(p,q) = 0$ when $p, q$ are sufficiently separated, therefore $H_p$ represents the local energy density around site $p$.

Following Ref. (1), the canonical choice for the energy current operator is

$$\begin{aligned} J_{pq}^E &= -i[H_p, H_q] \\ &= \frac{1}{8} \left( \zeta \cdot_p h \cdot_p J \cdot_q h \cdot_q \zeta - \zeta \cdot_q h \cdot_q J \cdot_p h \cdot_p \zeta \right). \end{aligned} \qquad [S51]$$

The first term in the parenthesis is an abbreviation of four different terms. The four terms correspond to four ways of inserting one $p$ and one $q$ into the dotted slots, and there are two choices for $p$ and two choices for $q$. For example, a term of the form $\zeta p h J q h \zeta$ means

$$\zeta p h J q h \zeta = \sum_{m,n} \zeta_p^I h^{IJ}(p,m) J^{JK}(m,q) h^{KL}(q,n) \zeta_n^L.$$

That is, at the insertions, the site indices are fixed to be $p, q$ respectively, and all other indices are contracted.

In later computations, we need to compute the contraction of the chain $J_{pq}^E$ with some coahins. The two contractions we need is

$$J^{E0}(\delta\alpha) \equiv \frac{1}{2} \sum_{pq} J_{pq}^E(\alpha(q) - \alpha(p)) = \frac{1}{8} \zeta \left(2[hJh, \alpha] + h[J, \alpha]h\right) \zeta. \qquad [S52]$$

$$f J^E g - g J^E f \equiv \sum_{pq} J_{pq}^E(f(p)g(q) - g(p)f(q)) = \frac{1}{4} \zeta J^{-1} [J(hf + fh), J(gh + hg)] \zeta \qquad [S53]$$

In the final result, the index summations can be absorbed into contractions between matrices and vectors of site index and cartesian index, and $\alpha$ is understood as a diagonal matrix in both kinds of indices. We will further assume that $\zeta = (u, \pi)$ consists of canonical coordinates and momenta, such that $J$ is the standard symplectic form Eq. (S6), and therefore $J$ commutes with all one-point functions, in particular $[J, \alpha] = 0$ in Eq. Eq. (S52).

**B. Interaction Part.** Adding the defect Hamiltonian and the phonon-defect interaction is equivalent to modifying the Hamiltonian density at site $o$ by $H_o \to H_o + \delta H_o$, $\delta H_o = H_{def} + H_{ph-def}$. Again using $J^E_{pq} = -i[H_p, H_q]$ we find that $J^E_{pq} \to J^E_{pq} + \delta J^E_{pq}$ where

$$\delta J^E_{p,o} = -\delta J^E_{o,p} = \frac{\gamma}{2}\sum_m (\zeta_p h(p,m) J(m,o) \delta V) \quad (p \neq o). \tag{S54}$$

Here we have expanded $\delta V$ from a 3D vector to 6D vector by filling zeroes. The summation over cartesian indices has the usual matrix multiplication structure and is suppressed.

The relevant contractions are

$$\delta J^E(\delta \alpha) = \frac{\gamma}{2}\zeta[hJ, \alpha]\delta V. \tag{S55}$$

$$f\delta J^E g - g\delta J^E f = \gamma\zeta(fhJg - ghJf)\delta V. \tag{S56}$$

In the above contractions, the operator $\delta V$ is supported only on $o$.

## 4. Thermal Hall conductivity

With the preparations above, we now calculate the thermal Hall conductivity of the phonon-defect system. Since we have assumed that in absence of the coupling $\gamma$, the phonons are just the usual acoustic phonons in an isotropic crystal, we expect the thermal Hall effect to start at order $\gamma^2$. We will first compute the Kubo contribution and next the magnetization contribution.

**A. Kubo Contribution.** The Kubo contribution to thermal Hall conductance is given by

$$\kappa^{\text{Kubo}}(f,g) = \beta^2 \lim_{s\to 0} \int_0^\infty dt\, e^{-st} \langle\langle J^E(\delta f, t); J^E(\delta g)\rangle\rangle. \tag{S57}$$

We now express Eq. (S57) in terms of the current-current correlation function

$$\Pi_{EE}(f,g;\tau) = \langle J^E(f,\tau) J^E(g) \rangle_c, \tag{S58}$$

$$\Pi_{EE}(f,g;i\omega_n) = \int_0^\beta d\tau\, e^{i\omega_n \tau} \Pi_{EE}(f,g;\tau). \tag{S59}$$

Using spectral representation of $\Pi_{EE}$, we can show that the Kubo pairing is equivalent to

$$\lim_{s\to 0}\int_0^\infty dt\, e^{-st} \int_0^\beta d\tau\, \Pi_{EE}(\tau + it) = -i\Pi'_{EE,+}(0). \tag{S60}$$

Here $\Pi_{EE,+}(z) = \Pi_{EE}(i\omega_n \to z + i0)$. The Kubo thermal conductivity is then given by

$$\kappa^{\text{Kubo}}(f,g) = -i\beta \Pi'_{EE,+}(f,g;0). \tag{S61}$$

The current-current correlator $\Pi_{EE}$ can be calculated using Wick's theorem[*], and there are the following contributions

$$\Pi^{(0)}_{EE}(f,g;i\omega_n) = 2T\sum_{\Omega_n} \text{Tr}\left[\frac{[hJh,f]}{4}D(i\omega_n + i\Omega_n)\frac{[hJh,g]}{4}D(i\Omega_n)\right]. \tag{S62}$$

$$\Pi^{(1)}_{EE}(f,g;i\omega_n) = 2T\sum_{\Omega_n} \text{Tr}\left[\frac{[hJh,f]}{4}D(i\omega_n + i\Omega_n)\frac{\gamma[hJ,g]}{2}G_{V\zeta}(i\Omega_n)\right.$$
$$\left. + \frac{\gamma[hJ,f]}{2}G_{V\zeta}(i\Omega_n + i\omega_n)\frac{[hJh,g]}{4}D(i\Omega_n)\right] \tag{S63}$$

---

[*] Strictly speaking, Wick's theorem doesn't apply to the defect system. However at order $\gamma^2$ we only need two-point functions of the defect system, and the correlators do factorize according to Wick's theorem.

$$\Pi_{EE}^{(2)}(f, g; i\omega_n) = \frac{\gamma^2}{8} T \sum_{\Omega_n} \text{Tr}\left[[hJ, f]^T D(i\omega_n + i\Omega_n)[hJ, g]S(i\Omega_n)\right]$$

$$+ \frac{\gamma^2}{8} T \sum_{\Omega_n} \text{Tr}\left[[hJ, f]S(i\Omega_n + i\omega_n)[hJ, g]^T D(i\Omega_n)\right] \quad \text{[S64]}$$

$$+ \frac{\gamma^2}{4} T \sum_{\Omega_n} \text{Tr}\left[[hJ, f]G_{V\zeta}(i\omega_n + i\Omega_n)[hJ, g]G_{V\zeta}(i\Omega_n)\right].$$

Here the superscripts count the power of $\gamma$ coming from the energy current vertices. The trace is over both site indices and cartesian indices.

In the above expressions, we have ignored vertex corrections due to phonon-defect coupling (which appears at order $\gamma^4$) and dissipation effects. While the ignorance of the latter can't be rigorously justified, we assume its effect is to renormalize the phonon self-energy lifetime to the transport lifetime, and we account for this by interpreting $1/\Gamma_{\text{ph}}$ in the phonon Green's function as the transport lifetime. Because phonon number is not conserved, we don't expect vertex corrections to bring qualitative changes.

The Green's functions $D, S, G_{V\zeta}$ appearing above are interacting Green's function defined as the following (we suppressed lattice indices)

$$D^{IJ}(\tau) = -T_\tau \left\langle \zeta^I(\tau)\zeta^J(0) \right\rangle, \quad \text{[S65]}$$

$$G_{V\zeta}^{IJ}(\tau) = -T_\tau \left\langle \delta V^I(\tau)\zeta^J(0) \right\rangle, \quad \text{[S66]}$$

$$S^{IJ}(\tau) = -T_\tau \left\langle \delta V^I(\tau)\delta V^J(0) \right\rangle. \quad \text{[S67]}$$

Notice that $G_{V\zeta}$ is of order $\gamma$, so the last line of $\Pi^{(2)}$ can be droppped.

To extract the Hall effects, we need to further antisymmetrize with respect to $f \leftrightarrow g$.

**B. Magnetization Correction.** The magnetization correction is given by

$$\mu^E(\delta f \cup \delta g) = \frac{1}{6} \sum_{pqr} \mu_{pqr}^E (f(q) - f(p))(g(r) - g(q)),$$

$$\mu_{pqr}^E = -\beta \left[ \langle\langle dH_p; J_{qr}^E \rangle\rangle + \langle\langle dH_r; J_{pq}^E \rangle\rangle + \langle\langle dH_q; J_{rp}^E \rangle\rangle \right]. \quad \text{[S68]}$$

Here $dH_p = \frac{\partial H_p}{\partial \gamma}$. By simple algebra, we can express $\mu^E$ in terms of contractions

$$\mu^E(\delta f \cup \delta g) = -\frac{\beta}{4} \left[ 2\langle\langle dH(g); J^E(\delta f) \rangle\rangle - 2\langle\langle dH(f); J^E(\delta g) \rangle\rangle + \langle\langle dH; fJ^E g - gJ^E f \rangle\rangle \right], \quad \text{[S69]}$$

where $dH = \frac{\partial H}{\partial \gamma} = \zeta \delta V$, $dH(f) = \zeta f \delta V$. For the final result to be order $\gamma^2$, $\mu_E$ needs to be calculated to linear order in $\gamma$.

Expressing everything in terms of Green's function, we obtain

$$\mu^E(\delta f \cup \delta g) = -\frac{1}{4} T \sum_{\Omega_n} \text{Tr}\left[ (gG_{V\zeta}[hJh, f]D + \gamma g D[hJ, f]S) - (f \leftrightarrow g) \right.$$

$$+ \frac{1}{2} G_{V\zeta} J^{-1}[J(hf + fh), J(hg + gh)]D \quad \text{[S70]}$$

$$\left. + \gamma D(fhJg - ghJf)S \right].$$

Here all Green's functions have argument $i\Omega_n$.

9

**C. Evaluation.** We compute the Matsubara sums and analytically continue to real frequency, to obtain

$$\kappa^{\text{Kubo},(0)}(f,g) = \frac{\beta}{16\pi} \int dz\, n_B(z)\, \text{Tr}\, \big[[hJh,f](-D'_+)[hJh,g](D_+ - D_-) \\ - [hJh,f](D_+ - D_-)[hJh,g](-D'_-)\big]\,, \quad \text{[S71]}$$

$$\kappa^{\text{Kubo},(1)}(f,g) = \frac{\beta\gamma}{8\pi} \int dz\, n_B(z)\, \text{Tr}\, \Big[[hJh,f](-D'_+)[hJ,g](G_{V\zeta,+} - G_{V\zeta,-}) \\ - [hJh,f](D_+ - D_-)[hJ,g](-G'_{V\zeta,-}) + [hJ,f](-G'_{V\zeta,+})[hJh,g](D_+ - D_-) \\ - [hJ,f](G_{V\zeta,+} - G_{V\zeta,-})[hJh,g](-D'_-)\Big]\,, \quad \text{[S72]}$$

$$\kappa^{\text{Kubo},(2)}(f,g) = \frac{\beta\gamma^2}{16\pi} \int dz\, n_B(z)\, \text{Tr}\, \Big[[Jh,f](-D'_+)[hJ,g](S_+ - S_-) \\ - [Jh,f](D_+ - D_-)[hJ,g](-S'_-) + [hJ,f](-S'_+)[Jh,g](D_+ - D_-) \\ - [hJ,f](S_+ - S_-)[Jh,g](-D'_-)\Big]\,. \quad \text{[S73]}$$

$$\mu^E(\delta f \cup \delta g) = -\frac{1}{8\pi i} \int dz\, n_B(z)\, \text{Tr}\, \Big[\Big(\gamma D_+[h,f]D_+[hJ,g]S_+ \\ + G_{V\zeta,+}[hJh,f]D_+[h,g]D_+ + \frac{1}{2}G_{V\zeta,+}[h,f]J[h,g]D_+ \\ - (f \leftrightarrow g)\Big) - (+ \to -)\Big]\,. \quad \text{[S74]}$$

Here $D_\pm = D(i\Omega_n \to z \pm i0)$ and similarly for $G_{V\zeta}$ and $S$. Prime means derivative with respect to $z$.

Next, we should antisymmetrize with respect to $f,g$, and also transform the Green's functions into a more convenient form, using the following substitutions

$$\tilde{h} = iJh \quad \text{[S75]}$$
$$\tilde{D} = D(iJ)^{-1} \quad \text{[S76]}$$
$$\tilde{S} = iJS \quad \text{[S77]}$$
$$\tilde{G}_{V\zeta} = iJG_{V\zeta}(iJ)^{-1} \quad \text{[S78]}$$

We can now get rid of $J$ and obtain

$$\kappa_H^{\text{Kubo},(0)}(f,g) = -\frac{\beta}{32\pi} \int dz\, n_B(z)\, \text{Tr}\, \big[[\tilde{h}^2,f](-\tilde{D}'_+)[\tilde{h}^2,g](\tilde{D}_+ - \tilde{D}_-) \\ - [\tilde{h}^2,f](\tilde{D}_+ - \tilde{D}_-)[\tilde{h}^2,g](-\tilde{D}'_-)\big] - (f \leftrightarrow g)\,, \quad \text{[S79]}$$

$$\kappa_H^{\text{Kubo},(1)} = -\frac{\beta\gamma}{16\pi} \int dz\, n_B(z)\, \text{Tr}\, \Big[[\tilde{h}^2,f](-\tilde{D}'_+)[\tilde{h},g](\tilde{G}_{V\zeta,+} - \tilde{G}_{V\zeta,-}) \\ - [\tilde{h}^2,f](\tilde{D}_+ - \tilde{D}_-)[\tilde{h},g](-\tilde{G}'_{V\zeta,-}) + [\tilde{h},f](-\tilde{G}'_{V\zeta,+})[\tilde{h}^2,g](\tilde{D}_+ - \tilde{D}_-) \\ - [\tilde{h},f](\tilde{G}_{V\zeta,+} - \tilde{G}_{V\zeta,-})[\tilde{h}^2,g](-\tilde{D}'_-)\Big] - (f \leftrightarrow g)\,, \quad \text{[S80]}$$

10

$$\kappa_H^{\text{Kubo},(2)} = -\frac{\beta\gamma^2}{32\pi} \int \mathrm{d}z\, n_B(z)\, \mathrm{Tr}\, \Big[ [\tilde{h}, f](-\tilde{D}'_+)[\tilde{h}, g](\tilde{S}_+ - \tilde{S}_-)$$
$$- [\tilde{h}, f](\tilde{D}_+ - \tilde{D}_-)[\tilde{h}, g](-\tilde{S}'_-) + [\tilde{h}, f](-\tilde{S}'_+)[\tilde{h}, g](\tilde{D}_+ - \tilde{D}_-) \qquad [\text{S81}]$$
$$- [\tilde{h}, f](\tilde{S}_+ - \tilde{S}_-)[\tilde{h}, g](-\tilde{D}'_-) \Big] - (f \leftrightarrow g).$$

$$\mu^E(\delta f \cup \delta g) = \frac{1}{8\pi} \int \mathrm{d}z\, n_B(z)\, \mathrm{Tr}\, \Big[ \Big( \gamma \tilde{D}_+[\tilde{h}, f]\tilde{D}_+[\tilde{h}, g]\tilde{S}_+$$
$$+ \tilde{G}_{V\zeta,+}[\tilde{h}^2, f]\tilde{D}_+[\tilde{h}, g]\tilde{D}_+ + \frac{1}{2}\tilde{G}_{V\zeta,+}[\tilde{h}, f][\tilde{h}, g]\tilde{D}_+ \qquad [\text{S82}]$$
$$- (f \leftrightarrow g) \Big) - (+ \to -) \Big]$$

In the above equations, all Green's functions are written using the interacting Green's function, which are subject to perturbation expansion in $\gamma$ that we carry out now.

To second order in $\gamma$, we have

$$\tilde{D} = \tilde{D}_0 + \gamma^2 \underbrace{\tilde{D}_0 \tilde{S}_0 \tilde{D}_0}_{\tilde{K}_0}, \qquad [\text{S83}]$$

$$\tilde{G}_{V\zeta} = \gamma \underbrace{\tilde{S}_0 \tilde{D}_0}_{\tilde{G}_0}, \qquad [\text{S84}]$$

$$\tilde{S} = \tilde{S}_0, \qquad [\text{S85}]$$

Here $\tilde{D}_0$ and $\tilde{S}_0$ denote the free Green's function. All Green's function appearing below refer to the free Green's function, and for notational simplicity, we will now suppress the subscript 0.

The full thermal Hall conductance is given by the differential equation with respect to $\gamma$.

$$\mathrm{d}\kappa_H(f, g) = \mathrm{d}\kappa_H^{\text{Kubo}}(f, g) - 2\beta \mu^E(\delta f \cup \delta g), \qquad [\text{S86}]$$

and the boundary condition is that $\kappa_A(f, g) = 0$ when $\gamma = 0$. To order $\gamma^2$, we have

$$\kappa_H(f, g) = \kappa_H^{\text{Kubo}}(f, g) - \gamma \beta \mu^E(\delta f \cup \delta g). \qquad [\text{S87}]$$

The above expression contains several pieces which are all of order $\gamma^2$, explicitly written below:

$$\kappa_H^{\text{Kubo},(0)}(f, g) = -\frac{\beta\gamma^2}{32\pi} \int \mathrm{d}z\, n_B(z)\, \mathrm{Tr}\, \Big[ [\tilde{h}^2, f](-\tilde{D}'_+ - \tilde{D}'_-)[\tilde{h}^2, g](\tilde{K}_+ - \tilde{K}_-)$$
$$+ [\tilde{h}^2, f](-\tilde{K}'_+ - \tilde{K}'_-)[\tilde{h}^2, g](\tilde{D}_+ - \tilde{D}_-) \Big] - (f \leftrightarrow g). \qquad [\text{S88}]$$

$$\kappa_H^{\text{Kubo},(1)} = -\frac{\beta\gamma^2}{16\pi} \int \mathrm{d}z\, n_B(z)\, \mathrm{Tr}\, \Big[ [\tilde{h}^2, f](-\tilde{D}'_+ - \tilde{D}'_-)[\tilde{h}, g](\tilde{G}_+ - \tilde{G}_-)$$
$$+ [\tilde{h}, f](-\tilde{G}'_+ - \tilde{G}'_-)[\tilde{h}^2, g](\tilde{D}_+ - \tilde{D}_-) \Big] - (f \leftrightarrow g), \qquad [\text{S89}]$$

$$\kappa_H^{\text{Kubo},(2)} = -\frac{\beta\gamma^2}{32\pi} \int \mathrm{d}z\, n_B(z)\, \mathrm{Tr}\, \Big[ [\tilde{h}, f](-\tilde{D}'_+ - \tilde{D}'_-)[\tilde{h}, g](\tilde{S}_+ - \tilde{S}_-)$$
$$+ [\tilde{h}, f](-\tilde{S}'_+ - \tilde{S}'_-)[\tilde{h}, g](\tilde{D}_+ - \tilde{D}_-) \Big] - (f \leftrightarrow g). \qquad [\text{S90}]$$

$$-\beta\gamma\mu^E(\delta f \cup \delta g) = -\frac{\beta\gamma^2}{8\pi}\int \mathrm{d}z\, n_B(z)\,\mathrm{Tr}\left[\left(\tilde{G}_+[\tilde{h},f]\tilde{D}_+[\tilde{h},g]\right.\right.$$
$$\left.+ \tilde{K}_+[\tilde{h}^2,f]\tilde{D}_+[\tilde{h},g] + \frac{1}{2}\tilde{K}_+[\tilde{h},f][\tilde{h},g]\right.$$
$$\left.\left.- (f \leftrightarrow g)\right) - (+ \to -)\right] \quad [\text{S91}]$$

The evaluation of the above integrals are done using Mathematica. The intermediate results are quite complicated so we just describe the algorithm.

The above integrals Eq. Eq. (S88)-Eq. (S91) are written as a trace over the single-particle indices (site index + cartesian index). The trace over site indices can be transformed into a momentum sum, according to

$$\mathrm{Tr} \to L^d \int \frac{\mathrm{d}^d k}{(2\pi)^d} \mathrm{tr}\,. \quad [\text{S92}]$$

Here tr denote the trace over cartesian indices ($2d$ by $2d$). $L$ is the system size and $d=3$ is the spatial dimension. The phonon Green's function is given by Eq. (S29) with a $k$-dependent $h$, and the defect Green's function is given by Eq. (S38) with $k_1 = k_2 = k$. Notice that although momentum does not have to conserve at the defect, the momentum trace naturally sets all momenta in the Green's functions to be equal. To further simplify, we switch to the band basis where $\tilde{h}$ and $\tilde{D}_\pm$ become diagonal. This can be done by diagonalizing the matrix $\tilde{h}(k) = iJh(k)$:

$$\tilde{h}(k) = M(k)\mathcal{E}(k)M(k)^{-1}\,, \quad [\text{S93}]$$

where $\mathcal{E}(k)$ is a diagonal matrix with entries $\pm c_T k$ and $\pm c_L k$.

As discussed in the main text, the function $f(p) = -x(p)/L$ and $g(p) = -y(p)/L$. In the continuum limit they should be replaced by momentum derivatives, which reads

$$[\tilde{h},f] = \frac{i}{L}\frac{\partial \tilde{h}(k)}{\partial k_x}\,, \quad [\tilde{h},g] = \frac{i}{L}\frac{\partial \tilde{h}(k)}{\partial k_y}\,, \quad [\text{S94}]$$

and similarly for $[\tilde{h}^2, f]$ and $[\tilde{h}^2, g]$. We see that overall the thermal Hall conductance scales with system size as $L^{d-2}$, as expected, and by dividing out the $L^{d-2}$ factor we obtain the thermal Hall conductivity.

By conjugating every term in Eq. (S88)-Eq. (S91) by $M(k)$, we arrive at a form where the phonon Green's function $\tilde{D}_\pm$ is diagonal. The energy current vertices $M(k)^{-1}[\tilde{h},\cdot]M(k)$, $M(k)^{-1}[\tilde{h}^2,\cdot]M(k)$ and the defect Green's function will contain both diagonal and off-diagonal components. Notice that the energy current vertices have no matrix element between the two degenerate transverse bands.

Next, we compute the $z$ integral in Eq. (S88)-Eq. (S91). We first use `Apart[]` in Mathematica to decompose all rational functions in $z$ into simple fractions. All the $z$ integrals can be calculated using the formula

$$\int_{-\infty}^{\infty}\mathrm{d}z\, n_B(z)\frac{1}{(z-a)^n} = \frac{1}{(n-1)!}\partial_a^n\left[\frac{i\pi}{\beta a} - \psi^{(0)}\left(\frac{i\beta a}{2\pi}\right) + 2\pi i n_B(a)\theta(\mathrm{Im} a)\right]\,, \quad [\text{S95}]$$

where $\theta$ is the step function and $\psi^{(0)}$ is the digamma function. The derivative doesn't act on the step function. Here, the pole of $n_B(z)$ at $z=0$ is resolved with principal value. The formula can be derived using a rectangular integral between $\mathrm{Im} z = 0$ and $\mathrm{Im} z = 2\pi i/\beta$. Notice that when $n=1$ the integral is logarithmically divergent, and we regulate the integral by ignoring the contribution from the path $z = -\infty \to z = -\infty + 2\pi i/\beta$. In the actual evaluation, the whole integral is UV convergent, so the result is independent of regularization.

The last step is to evaluate the $k$-integral. We first perform the angular integral of $k$. The integral over the magnitude of $k$ can't be performed analytically, and it is UV divergent. We are interested in the most singular part in the $\Gamma_{\text{ph}} \to 0$ limit. According to discussion in the main text, this should come from the side-jump effects and we expect it to scale as $1/\Gamma_{\text{ph}}$. This $1/\Gamma_{\text{ph}}$ enhancement comes from integrating pairs of retarded and advanced phonon Green's function of the same band over $z$. The integrand can then be classified into three types

1. $1/\Gamma_{\text{ph}}$ enhanced resonance: These terms have the $1/\Gamma_{\text{ph}}$ prefactor, and have resonant denominators $1/(c_T k - \epsilon \pm i\Gamma_{\text{ph}})$ or $1/(c_T k - \epsilon \pm i\Gamma_{\text{ph}})$ with $\epsilon > 0$. This is the resonant side-jump effect. This type of integrals can be calculated using the Sokhotski–Plemelj theorem and its derivatives. It turns out that only the $\delta$-function piece of Sokhotski–Plemelj theorem contributes, and the principal value parts cancelled out.

2. $1/\Gamma_{\text{ph}}$ enhanced non-resonant terms: These terms have the $1/\Gamma_{\text{ph}}$ prefactor, but doesn't have resonance at positive $k$. It can be shown explicitly that such terms are purely imaginary in the $\Gamma_{\text{ph}} \to 0$ limit, but it can be checked numerically that the total integrand is real, so these terms don't contribute

3. The rest terms are not enhanced by $1/\Gamma_{\text{ph}}$. These terms classified as the intrinsic contribution to thermal Hall conductivity. There is no simple formula to evaluate these terms and some of them are UV divergent. However, because of the absence of $1/\Gamma_{\text{ph}}$ factor, we expect these terms are subdominant to the side-jump effects.

Therefore, the resonant side-jump effect dominates the thermal Hall conductivity, and we obtain

$$\kappa_H^{sj} = \frac{1}{N_{\text{sys}}} \frac{\alpha_R \beta^2 \gamma^2 \Delta_Z m \epsilon_0^3 \left(e^{\beta\epsilon_0}(\beta\epsilon_0 - 4) + 4\right)\left(c_L^{-1} + c_T^{-1}\right)}{12\pi \Gamma_{\text{ph}} \left(e^{\beta\epsilon_0} + 3\right) \sinh^2(\beta\Delta/2)} \ . \quad [S96]$$

Some remarks:

1. Despite that the full integral is UV divergent, the contribution from the resonant peaks is UV finite, and it vanishes exponentially at low temperature, in agreement with the consistency requirement of thermal Hall conductance in Ref. (1).

2. By attaching a labelling variable to the diagonal parts of the vertex functions, we can show that this result exactly corresponds to the case where one vertex is intra-band and the other one is inter-band. The result is also enhanced by the phonon lifetime by $1/\Gamma_{\text{ph}}$. Both features are consistent with the side-jump effects studied in Ref. (2).

3. Again by using labelling variables, we can show that only Eq. (S88) and Eq. (S89) contribute to the final result. In particular the magnetization correction doesn't contribute. This can be understood as the following: The $1/\Gamma_{\text{ph}}$ enhancement essentially arises from a product of retarded and advanced phonon Green's function $D_+ D_-$ of the same band. In Eq. (S90), there is only one phonon Green's function. In Eq. (S91), the Green's functions are either all retarded or all advanced. This is an example showing that magnetization correction is unimportant for the extrinsic contributions.

4. The last factor $(c_L^{-1} + c_T^{-1})$ looks unnatural because there are two transverse bands but they only contribute only half the contribution of a longitudinal band. This is an artifact of using the isotropic phonon band structure, as the side-jump coordinate shift between two degenerate transverse bands should vanish due to gauge-invariance (see Sec.7). In a realistic model where the two transverse bands slightly split, we expect the answer should be $(c_L^{-1} + 2c_T^{-1})$.

13

## 5. Absence of Magnetization Correction in Extrinsic Effects

In this part we try to give an argument of why magnetization correction is unimportant for extrinsic effects. According to the diagrammatic study in Ref. (2), the $1/\Gamma_\text{ph}$ enhancement of extrinsic contributions (side jump or skew scattering) comes from a product of retarded and advanced phonon Green's function of the same band $(z - \omega_k + i\Gamma_\text{ph}/2)^{-1}(z - \omega_k - i\Gamma_\text{ph}/2)^{-1}$ that are attached to the energy current vertex, which after integrating over $z$ produces a factor of $1/\Gamma_\text{ph}$.

The magnetization correction, however, can't produce such a configuration of Green's functions. According to Eq. Eq. (S68), the correction term is a linear combination of Kubo correlation function, which is the $i\omega_n = 0$ case of a two-point function of bilinear operators, which has the generic form

$$\Pi(i\omega_n) \sim T \sum_{\Omega_n} D_1(i\omega_n + i\Omega_n) D_2(i\Omega_n) P(i\Omega_n, i\omega_n) D_3(i\omega_n + i\Omega_n) D_4(i\Omega_n) \,. \tag{S97}$$

Here we have only schematically written down the frequency sum and ignored other summations. In the complex $z = i\Omega_n$ plane, the integrand only has branch cut when $\text{Im} z = 0$ or $\text{Im} z = -i\omega_n$. Therefore, when $i\omega_n = 0$, all Green's functions are evaluated above the cut or below the cut, i.e. they are either all retarded or all advanced, and this is not a configuration for the $1/\Gamma_\text{ph}$ enhancement.

# Part II

# Models A and B

In this part we consider models A and B which have a more complicated structure in phonon-defect coupling. We will also derive a semiclassical expression for the thermal Hall effect.

## 6. General Phonon-Defect Coupling

The calculation of the first model assumes the phonon-defect coupling to be of a special inner product form. In this section we generalize the coupling to accommodate for more complex structure. The coupling between the phonon and the defect system can be written as

$$\delta H = \gamma \sum_q \zeta_q^K B_{qo}^{K\alpha} V_o^\alpha \,, \tag{S98}$$

where $V_o^\alpha$ is some operator of the defect, the coupling coefficients are encoded in $B$. Here $\gamma$ only serves as a power-counting parameter and will be set to one in the final result. We assume the matrix $B$ is translational invariant in its site indices, and therefore it has a well-defined fourier transform. We also assume $B_{qo}^{K\alpha}$ is quasi-diagonal, meaning that it vanishes when $q$ and $o$ are far apart.

Using the freedom of redefining Hamiltonian density (1), we can distribute all of $\delta H$ to $H_o$, and from this we can compute

$$\delta J^E_{po} = -\delta J^E_{op} = \frac{\gamma}{2} \left( \zeta^T h J p B V + \zeta^T p h J B V \right) \,, \tag{S99}$$

for $p \neq o$.

The contractions are then

$$\delta J^E(\delta\alpha) = \frac{\gamma}{2} \zeta^T \left( hJ[B, \alpha] + [hJB, \alpha] \right) V \,, \tag{S100}$$

$$f\delta J^E g - g\delta J^E f = \gamma \zeta^T \left( hJ(fBg - gBf) + (fhJBg - ghJBf) \right) V \,. \tag{S101}$$

To second order in phonon-defect coupling, the phonon Green's function is now

$$D = D^{(0)} + D^{(0)} \underbrace{\gamma^2 B S B^T}_{\Pi} D^{(0)}. \quad [S102]$$

We have also defined the phonon self energy $\Pi$. Note that in momentum space $(B^T)(k) = B(-k)^T$.

The crossed Green's function is

$$G = \gamma S B^T D. \quad [S103]$$

For notational simplicity we have dropped $V\zeta$ in the subscript. This expression is valid to all orders in $\gamma$ provided that $S$ and $D$ are the full defect Green's function and phonon Green's function respectively.

## 7. Semiclassical Expression for Phonon Thermal Hall Effect

In the computation of model C, we have shown that the side-jump thermal Hall effect comes from diagrams that exactly contains one pair of retarded and advanced phonon Green's function $D_+ D_-$ which has identical argument, and is therefore enhanced by a factor of $1/\Gamma_{ph}$. In this section we take advantage of this to derive a semiclassical expression for the thermal Hall effect. In what follows we set $\gamma = 1$.

Only the Kubo part contributes to side jump, and the current-current correlation functions we need are

$$\Pi_{EE}^{(0)}(f, g; i\omega_n) = 2T \sum_{\Omega_n} \text{Tr} \left[ \frac{[hJh, f]}{4} D(i\omega_n + i\Omega_n) \frac{[hJh, g]}{4} D(i\Omega_n) \right], \quad [S104]$$

$$\Pi_{EE}^{(1)}(f, g; i\omega_n) = 2T \sum_{\Omega_n} \text{Tr} \left[ \frac{[hJh, f]}{4} D(i\omega_n + i\Omega_n) \frac{1}{2} \left([hJB, g] + hJ[B, g]\right) G(i\Omega_n) \right.$$
$$\left. + \frac{1}{2} \left([hJB, f] + hJ[B, f]\right) G(i\Omega_n + i\omega_n) \frac{[hJh, g]}{4} D(i\Omega_n) \right]. \quad [S105]$$

We first compute $\Pi_{EE}^{(0)}$. Retain only terms with a pair of $D_+ D_-$, we obtain

$$\kappa_H^{(0)}(f, g) = \frac{\beta}{2\pi} \int dz (-n_B'(z)) \text{Tr} \left( \frac{[hJh, f]}{4} D_+(z) \frac{[hJh, g]}{4} D_-(z) \right) - (f \leftrightarrow g). \quad [S106]$$

To facilitate evaluation, we use change of variables slightly different from that in model C

$$D = M \tilde{D} M^{-1}(iJ) \quad [S107]$$
$$\Pi = (iJ)^{-1} M \tilde{\Pi} M^{-1}. \quad [S108]$$

Here $M = M(k)$ is a momentum-dependent matrix that diagonalizes the Hamiltonian

$$M^{-1}(k) i J h(k) M(k) = \mathcal{E}(k), \quad [S109]$$

where $\mathcal{E}(k)$ is a diagonal matrix in band indices which encodes the phonon dispersion. In this basis the self energy reads

$$\tilde{\Pi} = M^{-1} i J B S B^T M. \quad [S110]$$

Then Eq. Eq. (S106) becomes

$$\kappa_H^{(0)}(f, g) = \frac{-\beta}{2\pi} \int dz (-n_B'(z)) \text{Tr} \left[ V_f^{(0)} \tilde{D}_+ V_g^{(0)} \tilde{D}_- \right] - (f \leftrightarrow g), \quad [S111]$$

and the vertex function is

$$V_f^{(0)} = \frac{M^{-1}[(iJh)^2, f]M}{4} = \frac{1}{4}\left([\mathcal{E}^2, f] + [\mathcal{E}^2, A_f]\right),$$ [S112]

where we have separated it into band-diagonal and band-off diagonal parts. The off-diagonal part is given by the connection $A_f$:

$$A_f = -M^{-1}[M, f].$$ [S113]

We recall that the formula for converting commutator to derivative is $[H(k), f] = i\partial_{k_x}H(k)$, $[H(k), g] = i\partial_{k_y}H(k)$.

Now expanding Eq. Eq. (S111) to first order in self energy, we obtain

$$\begin{aligned}\kappa_H^{(0)} &= \frac{-\beta}{2\pi}\int \mathrm{d}z(-n_B'(z))\,\mathrm{Tr}\,[V_f^{(0)}\tilde{D}_+^{(0)}\tilde{\Pi}_+\tilde{D}_+^{(0)}V_g^{(0)}\tilde{D}_-^{(0)} + V_f^{(0)}\tilde{D}_+^{(0)}V_g^{(0)}\tilde{D}_-^{(0)}\tilde{\Pi}_-\tilde{D}_-^{(0)}] \\ &\quad - (f \leftrightarrow g)\,.\end{aligned}$$ [S114]

We proceed to pick out the side-jump contribution, using the formula

$$\tilde{D}_+^{(0),a}(z)\tilde{D}_-^{(0),a}(z) = \frac{2\pi}{\Gamma_a(z)}\delta(z - \mathcal{E}_a),$$ [S115]

where $a$ here denotes an eigenstate of the quadratic phonon Hamiltonian $iJh$ and $\Gamma_a$ is its decay rate.

We obtain

$$\begin{aligned}\kappa_H^{(0)} &= \sum_{a,b:\mathcal{E}_a \neq \mathcal{E}_b}\frac{\beta n_B'(\mathcal{E}_a)}{2\Gamma_a}\mathcal{E}_a[\mathcal{E}_a, f]\frac{\mathcal{E}_a + \mathcal{E}_b}{4}\left[\left(\tilde{\Pi}_-^{ab}(\mathcal{E}_a) - \tilde{\Pi}_+^{ab}(\mathcal{E}_a)\right)A_g^{ba} + \left(\tilde{\Pi}_-^{ba}(\mathcal{E}_a) - \tilde{\Pi}_+^{ba}(\mathcal{E}_a)\right)A_g^{ab}\right] \\ &\quad - (f \leftrightarrow g)\,.\end{aligned}$$ [S116]

Here the summation indices $a, b$ are general, which include both momentum and band indices. Here we notice that the off-diagonal component of the vertex function is $V_f^{(0),ab} = A_f^{ab}(\mathcal{E}_a^2 - \mathcal{E}_b^2)/4$, which vanishes when the two bands are degenerate, and therefore the summation should be restricted to bands with distinct energy.

The above result can be simplified by noting that the two terms in the bracket are actually equal. To show this we need the transpose properties of the connection $A_f$ and the self energy $\tilde{\Pi}$, which are

$$A_f^T = -K^{-1}A_f K,$$ [S117]

$$\tilde{\Pi}_\pm^T(z) = -K^{-1}\tilde{\Pi}_\mp(-z)K,$$ [S118]

where $K$ is defined in Eq. (S20). Eq. (S117) follows from its definition Eq. (S113) and Eq. (S118) can be derived from the symmetry of the Green's function $D^{ab}(\tau) \equiv -T_\tau \left\langle \zeta^a(\tau)\zeta^b(0)\right\rangle = D^{ba}(-\tau)$ under $a \leftrightarrow b$. Following the discussion in the section of phonon diagonalization A, we know that the matrix $K$ only has matrix element $K_{a\bar{a}} = \pm 1$ between mode $a$ and mode $\bar{a}$ of opposite eigenvalue $\mathcal{E}_{\bar{a}} = -\mathcal{E}_a$. We can therefore eliminate $K$ by changing the summation variables to $\bar{a}, \bar{b}$ for the second term in the bracket, which flips the sign of $\mathcal{E}$ in the summand, and the result is equal to the first term.

We proceed by writing out the momentum summation explicitly:

$$\begin{aligned}\kappa_H^{(0)} &= \sum_{a,b:\mathcal{E}_a \neq \mathcal{E}_b}\int\frac{\mathrm{d}^3k}{(2\pi)^3}\frac{\beta n_B'(\mathcal{E}_a(k))}{2\Gamma_a(k)}\mathcal{E}_a(k)[\mathcal{E}_a(k), f]A_g^{ba}(k)\frac{\mathcal{E}_a(k) + \mathcal{E}_b(k)}{2}\left(\tilde{\Pi}_-^{ab}(\mathcal{E}_a(k)) - \tilde{\Pi}_+^{ab}(\mathcal{E}_a(k))\right) \\ &\quad - (f \leftrightarrow g)\,.\end{aligned}$$ [S119]

Now the indices $a, b$ are only over band indices. The commutator $[\mathcal{E}_a(k), f]$ is equivalent to $i\partial_{k_x}\mathcal{E}_a(k)$ and the connection $A_g$ is $A_g(k) = -iM(k)^{-1}\partial_{k_y}M(k)$.

16

Next we evaluate the contribution from $\Pi_{EE}^{(1)}$ which can be decomposed as $\kappa_H^{(1)} = \kappa_H^{(1a)} + \kappa_H^{(1b)}$:

$$\kappa_H^{(1a)} = \frac{\beta}{2\pi} \int \mathrm{d}z (-n_B'(z)) \operatorname{Tr}\left[ \frac{[hJh,f]}{4} D_+ \frac{[h,g]JB}{2} G_- + \frac{[h,f]JB}{2} G_+ \frac{[hJh,g]}{4} D_- \right] - (f \leftrightarrow g), \quad [\text{S120}]$$

$$\kappa_H^{(1b)} = \frac{\beta}{2\pi} \int \mathrm{d}z (-n_B'(z)) \operatorname{Tr}\left[ \frac{[hJh,f]}{4} D_+ h[JB,g] G_- + h[JB,f] G_+ \frac{[hJh,g]}{4} D_- \right] - (f \leftrightarrow g). \quad [\text{S121}]$$

To make contact of $\kappa_H^{(1a)}$ to $\kappa_H^{(0)}$, we substitute Eq. (S103) into Eq. (S120) and take transpose within the trace:

$$\kappa_H^{(1a)} = \frac{\beta}{2\pi} \int \mathrm{d}z (-n_B'(z)) \operatorname{Tr}\left[ \frac{[hJh,f]}{4} D_+ BS_+ \frac{B^T J[h,g]}{2} D_- + \frac{B^T J[h,f]}{2} D_+ \frac{[hJh,g]}{4} D_- BS_- \right] \quad [\text{S122}]$$
$$- (f \leftrightarrow g),$$

where we have used the fact that $D_\pm(z)^T = D_\mp(-z), S_\pm(z)^T = S_\mp(-z), n_B'(-z) = n_B'(z)$. Using the expression for phonon self energy, we have

$$\kappa_H^{(1a)} = \sum_{a,b:\mathcal{E}_a \neq \mathcal{E}_b} \int \frac{\mathrm{d}^3 k}{(2\pi)^3} \frac{\beta n_B'(\mathcal{E}_a(k))}{2\Gamma_a(k)} \mathcal{E}_a(k) [\mathcal{E}_a(k), f] A_g^{ba}(k) \frac{\mathcal{E}_a(k) - \mathcal{E}_b(k)}{2} \left( \tilde{\Pi}_-^{ab}(\mathcal{E}_a(k)) - \tilde{\Pi}_+^{ab}(\mathcal{E}_a(k)) \right) \quad [\text{S123}]$$
$$- (f \leftrightarrow g).$$

The other term $\kappa_H^{(1b)}$ is

$$\kappa_H^{(1b)} = \sum_a \int \frac{\mathrm{d}^3 k}{(2\pi)^3} \frac{\beta n_B'(\mathcal{E}_a(k))}{2\Gamma_a(k)} \mathcal{E}_a(k)^2 [\mathcal{E}_a(k), f] \left( M^{-1}[iJB,g](S_-(\mathcal{E}_a) - S_+(\mathcal{E}_a)) B^T M \right)^{aa} \quad [\text{S124}]$$
$$- (f \leftrightarrow g).$$

The total side-jump thermal Hall effect is therefore

$$\kappa_H^{sj} = \kappa_H^{(0)} + \kappa_H^{(1a)} + \kappa_H^{(1b)}. \quad [\text{S125}]$$

This result has a very clear semiclassical interpretation. It can be rewritten into a form similar to the solution of a semiclassical Boltzmann equation:

$$\kappa_H^{sj} = \frac{1}{2} \sum_a \int \frac{\mathrm{d}^3 k}{(2\pi)^3} \frac{(-\beta n_B'(\mathcal{E}_a))}{\Gamma_a(k)} j_{\text{on-shell}}^E(f) j_{\text{side-jump}}^E(g) - (f \leftrightarrow g). \quad [\text{S126}]$$

Here $j_{\text{on-shell}}^E$ denotes the energy current of an on-shell phonon, which takes the form

$$j_{\text{on-shell}}^E(f) = -i\mathcal{E}_a [\mathcal{E}_a, f]. \quad [\text{S127}]$$

Converting the commutator to momentum derivative, this is exactly energy times velocity. The other part is the side-jump energy current, which is

$$j_{\text{side-jump}}^E(g) = (v_{\text{sj}}^{aa}(g) + \sum_{b:\mathcal{E}_b \neq \mathcal{E}_a} v_{\text{sj}}^{ba}(g)) \mathcal{E}_a, \quad [\text{S128}]$$

and the side-jump velocities are

$$v_{\text{sj}}^{aa}(g) = \left( M^{-1}[iJB,g](-i)(S_-(\mathcal{E}_a) - S_+(\mathcal{E}_a)) B^T M \right)^{aa}, \quad [\text{S129}]$$

$$v_{\text{sj}}^{ba}(g) = A_g^{ba}(-i) \left( \tilde{\Pi}_-^{ab}(\mathcal{E}_a) - \tilde{\Pi}_+^{ab}(\mathcal{E}_a) \right). \quad [\text{S130}]$$

In the above equations, both $M$ and $\mathcal{E}_a$ are functions of the phonon momentum $k$. The phonon self energy $\tilde{\Pi}$ and spin Green's function are evaluated at $z = \mathcal{E}_a$. The expression for $\kappa_H$ in the main text is obtained by converting commutators to derivatives according to Eq. (S94).

Several remarks are in order:

1. Eq. (S126) takes the form as a solution to a semiclassical Boltzmann equation: $-\beta n'_B(\mathcal{E}_a)$ is the boson thermal weight, and $1/\Gamma_a$ is the phonon lifetime. Because our phonon band contains both positive and negative energy modes, there is an overall $1/2$ factor in the front. There are two energy currents $j^E_{\text{on-shell}}$ and $j^E_{\text{side-jump}}$: One couples to the temperature gradient and the other one is measured as an observable, and the two configurations are both included via the $f \leftrightarrow g$ antisymmetrization.

2. The first side-jump velocity $v^{aa}_{\text{sj}}$ occurs when the phonon-defect coupling $B$ has momentum dependence, which reflects the fact that a momentum-dependent coupling $B$ renormalizes phonon velocity.

3. The second side-jump velocity $v^{ba}_{\text{sj}}$ appears for arbitrary phonon-defect coupling. It appears as a product of the coordinate shift $A^{ba}_g$ and the scattering rate $(-i)\left(\tilde{\Pi}^{ab}_-(\mathcal{E}_a) - \tilde{\Pi}^{ab}_+(\mathcal{E}_a)\right)$. The coordinate shift we have derived is the inter-band berry connection Eq. (S113) which is automatically gauge invariant for $a \neq b$, while the semiclassical result in Ref. (3) is the difference of intra-band berry connections plus the argument of T-matrix to restore gauge invariance.

4. The coordinate shift $A^{ba}_f(k)$ is gauge invariant when $\mathcal{E}_a(k) \neq \mathcal{E}_b(k)$. The gauge degrees freedom of the diagonalization matrix $M(k)$ is $M(k) \to M(k)U(k)$ where $[U(k), \mathcal{E}(k)] = 0$. This condition requires $U(k)$ to be a block-diagonal matrix with nonzero elements only between degenerate bands. The corresponding change of $A^{ab}_f(k)$ will be $A^{ab}_f(k) \to \left(U(k)^{-1} A_f(k) U(k)\right)^{ab} - \left(U^{-1}(k)[U(k), f]\right)^{ab}$ where the commutator is proportional to some momentum derivative. Therefore, the change of $A^{ab}_f(k)$ will also be block-diagonal, meaning the components between non-degenerate bands are non-zero. However, components of $A^{ab}_f(k)$ between degenerate bands are not gauge-invariant, but it has been naturally excluded in the calculation.

5. The inclusion of the phonon-spin coupling into energy current (i.e. including $\Pi^{(1)}_{EE}$) is crucial for energy conservation. Without this term, the side-jump energy current will look like

$$j^{E,(0)}_{\text{side-jump}}(g) = \sum_{b \neq a} v^{ba}_{\text{sj}}(g) \frac{\mathcal{E}_b + \mathcal{E}_a}{2} .$$

The energy transported by the phonon during the jump is wrong.

6. Since one of the phonon in the side jump is off shell, one may wonder whether high energy optical phonons can contribute. The answer is no for the following reason: Assuming the acoustic branches are non topological (zero total Chern number), the Hamiltonian can then be written in a block diagonal form $H = H_{acoustic} \oplus H_{optical}$. The coordinate shift $A^{ab}_f(k)$ between two bands is proportional to $M^{-1}(k)\partial_{k_x} M(k)$ and $M(k)$ must also take a block diagonal form $M(k) = M_{acoustic}(k) \oplus M_{optical}(k)$, implying that $A_f$ between the acoustic and the optical branch is zero.

## 8. Application to the Models A and B

Now we apply the above formalism to the first two models in the main text. For model A, we have

$$B^{I\alpha}(k) = K_{i\alpha} , \qquad [\text{S131}]$$

where $i$ means momentum components.

In model B the continuum limit of $B$ should be

$$B^{I\alpha}(k) = -ik^l K_{li\alpha} . \qquad [\text{S132}]$$

where now $i$ denotes displacement components.

In both models, the spin-spin correlation function is

$$S^{\alpha\beta}_{\pm}(z) = 2\tanh\left(\frac{\beta\Delta}{2}\right)\begin{pmatrix} \frac{\Delta}{z_{\pm}^2-\Delta^2} & \frac{iz}{z_{\pm}^2-\Delta^2} & 0 \\ -\frac{iz}{z_{\pm}^2-\Delta^2} & \frac{\Delta}{z_{\pm}^2-\Delta^2} & 0 \\ 0 & 0 & 0 \end{pmatrix},\qquad [\text{S133}]$$

where $z_{\pm} = z \pm i0$.

The integrals in Eq. (S119), Eq. (S123), Eq. (S124) are easy to evaluate since they are all proportional to a delta function at resonance. We obtain for model A

$$\kappa_H^{sj} = \frac{1}{N_{\text{sys}}}\frac{\beta^2\Delta^4 m(c_L^{-1}+c_T^{-1})\text{csch}(\beta\Delta)}{6\pi\Gamma_{\text{ph}}}\left(K_{x1}K_{y2} - K_{x2}K_{y1}\right),\qquad [\text{S134}]$$

The last factor of Eq. (S134) can be understood from symmetry considerations: The thermal Hall conductivity $\kappa_{xy}$ is invariant under spatial $SO(2)_z$ along the z axis, and it is odd under spatial reflections $R_x, R_y$. These two conditions require a quadratic combination of $K_{i\alpha}$ with exactly one $x$ and one $y$ index, and is invariant under $SO(2)_z$, yielding $K_{x\alpha}K_{x\beta} + K_{y\alpha}K_{y\beta}$ or $K_{x\alpha}K_{y\beta} - K_{y\alpha}K_{x\beta}$. Next, notice that the choice of the spin frame has a redundancy because we only fix the '3' direction to align with the polarization of the defect, and then the remaining '1' and '2' axes can rotate freely around the '3' axis. Physical observables should be independent of this free rotating angle, and only the coupling combination $K_{x1}K_{y2} - K_{x2}K_{y1}$ satisfies the requirement.

For model B, we obtain

$$\kappa_H^{sj} = \frac{1}{N_{\text{sys}}}\frac{\beta^2\Delta^4\text{csch}(\beta\Delta)}{30\pi\Gamma_{\text{ph}}m}\left(c_L^{-3}K_L + c_T^{-3}K_T\right),\qquad [\text{S135}]$$

where

$$\begin{aligned}
K_L =\ & +2\left(K_{xy2}+K_{yx2}\right)\left(K_{xx1}-K_{yy1}\right) - 2\left(K_{xy1}+K_{yx1}\right)\left(K_{xx2}-K_{yy2}\right) \\
& + K_{zx1}K_{zy2} - K_{zx2}K_{zy1} \\
& + K_{xz1}K_{yz2} - K_{xz2}K_{yz1} \\
& + K_{zx1}K_{yz2} - K_{zx2}K_{yz1} - K_{xz2}K_{zy1} + K_{xz1}K_{zy2},
\end{aligned}\qquad [\text{S136}]$$

$$\begin{aligned}
K_T =\ & -\frac{5}{2}\left(\left(K_{xy1}+K_{yx1}\right)\left(K_{xx2}-K_{yy2}\right) - \left(K_{xy2}+K_{yx2}\right)\left(K_{xx1}-K_{yy1}\right)\right) \\
& + \frac{1}{2}\left(\left(K_{xy1}-K_{yx1}\right)\left(K_{xx2}+K_{yy2}\right) - \left(K_{xy2}-K_{yx2}\right)\left(K_{xx1}+K_{yy1}\right)\right) \\
& + K_{zx1}K_{zy2} - K_{zx2}K_{zy1} \\
& + 4K_{xz1}K_{yz2} - 4K_{xz2}K_{yz1} \\
& + K_{zz1}\left(K_{xy2}-K_{yx2}\right) + K_{zz2}\left(K_{yx1}-K_{xy1}\right).
\end{aligned}\qquad [\text{S137}]$$

The above result for model B has several features: First, same as model A, the spin-spin correlation function requires the coupling constants appear in antisymmetric combinations under $1 \leftrightarrow 2$, and the 3 component is absent. Second, the invariance of $\kappa_{xy}$ under $SO(2)_z$ and its covariance under $R_x$, $R_y$ require the couplings form $SO(2)_z$ invariant combinations with odd number of $x$ and $y$ indices. Each line of Eq. (S136) and Eq. (S137) satisfies the above conditions.

19

## 9. Model B from a lattice antiferromagnet

Let us consider a coupled-ladder Heisenberg antiferromagnet in a transverse field (see Fig. (S1)), thereby leading to canting of the moments. There is also a vacancy leading to a spin defect. We have the following Hamiltonian:

$$H = \sum_{\langle pq \rangle \in \mathcal{D}} J_{pq} \vec{S}_p \cdot \vec{S}_q + \sum_{\langle pq \rangle \notin \mathcal{D}} K_{pq}^\alpha S_p^\alpha S_q^\alpha + h \sum_p S_p^3. \qquad [S138]$$

Here $\mathcal{D}$ denotes a dimer unit cell (see Fig. (S1)). The real-space cartesian coordinates are $x, y, z$, while the spin-space indices are labelled as $1, 2, 3$. For the inter-dimer couplings we have considered an anisotropic exchange both in the spin space as well as the real space. The exchange interaction along the $\hat{x}$−direction is $K'^\alpha$, while that in the $\hat{y}$−direction is $K^\alpha$. We are interested in spin-phonon interaction. Such an interaction has been considered earlier in the context of magnetization plataues in quantum magnets (4). Here our interest is to study the effect of such interactions on thermal-Hall conductivity. Spin-phonon coupling arises naturally if we consider that the spin-exchange interactions depend on the bond distances, $J_{pq} \equiv J(|\vec{R}_p - \vec{R}_q|)$ and $K_{pq} \equiv K(|\vec{R}_p - \vec{R}_q|)$. Let $\vec{u}_p$ be the small displacement of the lattice site at $\vec{R}_p$ from its equilibrium position. We can then expand the coupling constants in terms of $\vec{u}_p$ to linear order,

$$J_{pq} = J(|\vec{R}_p - \vec{R}_q|) + \frac{dJ}{dR}\hat{x} \cdot (\vec{u}_p - \vec{u}_q) \equiv J\left(1 + \alpha \hat{x} \cdot (\vec{u}_p - \vec{u}_q)\right), \qquad [S139]$$

$$K_{pq}^\alpha = K^\alpha(|\vec{R}_p - \vec{R}_q|) + \frac{dK^\alpha}{dR}\hat{e}_{pq} \cdot (\vec{u}_p - \vec{u}_q) \equiv K^\alpha\left(1 + \beta^\alpha \hat{e}_{pq} \cdot (\vec{u}_p - \vec{u}_q)\right), \qquad [S140]$$

$$K_{pq}'^\alpha = K'^\alpha(|\vec{R}_p - \vec{R}_q|) + \frac{dK'^\alpha}{dR}\hat{e}_{pq} \cdot (\vec{u}_p - \vec{u}_q) \equiv K'^\alpha\left(1 + \beta'^\alpha \hat{e}_{pq} \cdot (\vec{u}_p - \vec{u}_q)\right). \qquad [S141]$$

We will label the defect site by the index $o$. Using the notation, $u_o^j - u_{o\pm\hat{i}}^j = \pm \partial_i u^j$, with $i, j = x, y, z$, the defect spin Hamiltonian is

$$\begin{aligned} H_o = &\sum_{\alpha=1,2,3} \left[ K'^\alpha \langle S_{o+\hat{x}}^\alpha \rangle S_o^\alpha + K^\alpha \left( \langle S_{o+\hat{y}}^\alpha \rangle + \langle S_{o-\hat{y}}^\alpha \rangle \right) S_o^\alpha \right] + h S_o^3 \\ &+ K'^1 \beta'^1 \langle S_{o+\hat{x}}^1 \rangle \partial_x u^x S_o^1 + K'^2 \beta'^2 \langle S_{o+\hat{x}}^2 \rangle \partial_x u^x S_o^2 + K'^3 \beta'^3 \langle S_{o+\hat{x}}^3 \rangle \partial_x u^x S_o^3 \\ &+ K^1 \beta^1 \left( \langle S_{o+\hat{y}}^1 \rangle + \langle S_{o-\hat{y}}^1 \rangle \right) \partial_y u^y S_o^1 + K^2 \beta^2 \left( \langle S_{o+\hat{y}}^2 \rangle + \langle S_{o-\hat{y}}^2 \rangle \right) \partial_y u^y S_o^2 \\ &+ K^3 \beta^3 \left( \langle S_{o+\hat{y}}^3 \rangle + \langle S_{o-\hat{y}}^3 \rangle \right) \partial_y u^y S_o^3. \end{aligned} \qquad [S142]$$

We identify the coefficient of the term $\partial_i u^j S_o^\alpha$ by $K_{ij\alpha}$, with $ij = x, y, z$ and $\alpha = 1, 2, 3$. Thus we have,

$$\begin{aligned} K_{xx1} &= K'^1 \beta'^1 \langle S_{o+\hat{x}}^1 \rangle, \quad K_{xx2} = K'^2 \beta'^2 \langle S_{o+\hat{x}}^2 \rangle, \quad K_{xx3} = K'^3 \beta'^3 \langle S_{o+\hat{x}}^3 \rangle \\ K_{yy1} &= K^1 \beta^1 \left( \langle S_{o+\hat{y}}^1 \rangle + \langle S_{o-\hat{y}}^1 \rangle \right), \quad K_{yy2} = K^2 \beta^2 \left( \langle S_{o+\hat{y}}^2 \rangle + \langle S_{o-\hat{y}}^2 \rangle \right), \quad K_{yy3} = K^3 \beta^3 \left( \langle S_{o+\hat{y}}^3 \rangle + \langle S_{o-\hat{y}}^3 \rangle \right). \end{aligned} \qquad [S143]$$

**A. Next-nearest neighbors.** Let us also consider the next-nearest neighbors, which are coupled to the defect spin by a coupling constant $K''$. Just as above,

$$K''^\alpha_{o,o+\hat{x}+\hat{y}} = K''^\alpha \left[ 1 + \beta''^\alpha (\hat{x}+\hat{y}) \cdot (\vec{u}_o - \vec{u}_{o+\hat{x}+\hat{y}}) \right]$$
$$= K''^\alpha \left[ 1 + \beta''^\alpha (\partial_x u^x + \partial_y u^x + \partial_x u^y + \partial_y u^y) \right] \quad [S144]$$

$$K''^\alpha_{o,o+\hat{x}-\hat{y}} = K''^\alpha \left[ 1 + \beta''^\alpha (\hat{x}-\hat{y}) \cdot (\vec{u}_o - \vec{u}_{o+\hat{x}-\hat{y}}) \right]$$
$$= K''^\alpha \left[ 1 + \beta''^\alpha (\partial_x u^x - \partial_y u^x - \partial_x u^y + \partial_y u^y) \right] \quad [S145]$$

$$K''^\alpha_{o,o-(\hat{x}+\hat{y})} = K''^\alpha \left[ 1 - \beta''^\alpha (\hat{x}+\hat{y}) \cdot (\vec{u}_o - \vec{u}_{o-(\hat{x}+\hat{y})}) \right]$$
$$= K''^\alpha \left[ 1 + \beta''^\alpha (\partial_x u^x + \partial_y u^x + \partial_x u^y + \partial_y u^y) \right] \quad [S146]$$

$$K''^\alpha_{o,o-\hat{x}+\hat{y}} = K''^\alpha \left[ 1 - \beta''^\alpha (\hat{x}-\hat{y}) \cdot (\vec{u}_o - \vec{u}_{o-\hat{x}+\hat{y}}) \right]$$
$$= K''^\alpha \left[ 1 + \beta''^\alpha (\partial_x u^x - \partial_y u^x - \partial_x u^y + \partial_y u^y) \right]. \quad [S147]$$

We therefore have the following additional terms in the defect-spin Hamiltonian:

$$H_{o,2n} = \sum_{\alpha=1,2,3} K''^\alpha P^\alpha S_o^\alpha$$
$$+ \sum_{\alpha=1,2,3} K''^\alpha \beta''^\alpha P^\alpha (\partial_x u^x + \partial_y u^y) S_o^\alpha + \sum_{\alpha=1,2,3} K''^\alpha \beta''^\alpha M^\alpha (\partial_x u^y + \partial_y u^x) S_o^\alpha, \quad [S148]$$

where

$$P^\alpha = \sum_{i=\pm x, j=\pm y} \langle S^\alpha_{o+\hat{i}+\hat{j}} \rangle, \quad M^\alpha = \sum_{\eta=\pm 1} \langle S^\alpha_{o+\eta(\hat{x}+\hat{y})} \rangle - \sum_{\eta=\pm 1} \langle S^\alpha_{o+\eta(\hat{x}-\hat{y})} \rangle. \quad [S149]$$

In terms of the convention introduced above we have the following couplings:

$$K_{xx1} = K_{yy1} = K''^1 \beta''^1 P^1, \quad K_{xx2} = K_{yy2} = K''^2 \beta''^2 P^2, \quad K_{xx3} = K_{yy3} = K''^3 \beta''^3 P^3$$
$$K_{xy1} = K_{yx1} = K''^1 \beta''^1 M^1, \quad K_{xy2} = K_{yx2} = K''^2 \beta''^2 M^2, \quad K_{xy3} = K_{yx3} = K''^3 \beta''^3 M^3. \quad [S150]$$

The thermal-Hall conductivity is proportional to terms like $(K_{xx\alpha} - K_{yy\alpha})(K_{xy\beta} + K_{yx\beta})$ and $(K_{xx\alpha} + K_{yy\alpha})(K_{xy\beta} - K_{yx\beta})$. If we consider only nearest neighbors then we can not generate terms like $K_{xy\alpha}$ etc. Hence the thermal-Hall conductivity vanishes trivially for any spin ordering. Whereas if we consider only second nearest neighbors we always have $K_{xx\alpha} = K_{yy\alpha}$ as well as $K_{xy\alpha} = K_{yx\alpha}$. Thus, there is no thermal-Hall in this case as well, irrespective of the spin ordering. However, if we have both first neighbor and the second neighbor interactions then

$$K_{xx1} = K'^1 \beta'^1 \langle S^1_{o+\hat{x}} \rangle + K''^1 \beta''^1 P^1, \quad K_{xx2} = K'^2 \beta'^2 \langle S^2_{o+\hat{x}} \rangle + K''^2 \beta''^2 P^2,$$
$$K_{xx3} = K'^3 \beta'^3 \langle S^3_{o+\hat{x}} \rangle + K''^3 \beta''^3 P^3$$
$$K_{yy1} = K^1 \beta^1 \left( \langle S^1_{o+\hat{y}} \rangle + \langle S^1_{o-\hat{y}} \rangle \right) + K''^1 \beta''^1 P^1, \quad K_{yy2} = K^2 \beta^2 \left( \langle S^2_{o+\hat{y}} \rangle + \langle S^2_{o-\hat{y}} \rangle \right) + K''^2 \beta''^2 P^2,$$
$$K_{yy3} = K^3 \beta^3 \left( \langle S^3_{o+\hat{y}} \rangle + \langle S^3_{o-\hat{y}} \rangle \right) + K''^3 \beta''^3 P^3,$$
$$K_{xy1} = K_{yx1} = K''^1 \beta''^1 M^1, \quad K_{xy2} = K_{yx2} = K''^2 \beta''^2 M^2, \quad K_{xy3} = K_{yx3} = K''^3 \beta''^3 M^3. \quad [S151]$$

In this case, for a non-trivial spin ordering we may have a non-zero thermal-Hall conductivity. In particular, for a spin-glass order we may generically obtain a non-zero thermal-Hall. Additionally, this will most likely be independent

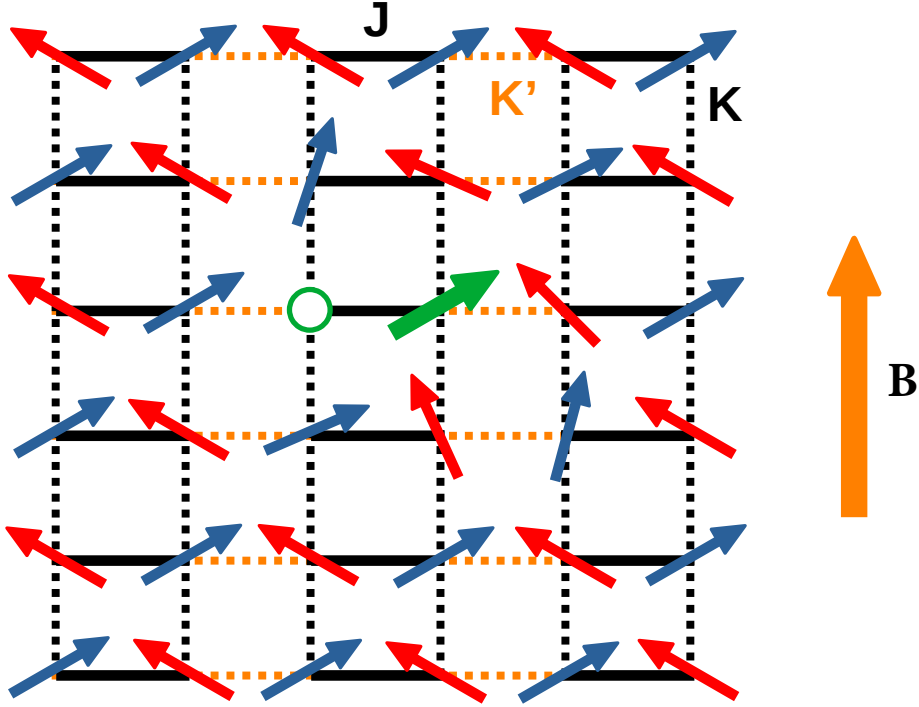

**Fig. S1.** Solid lines, i.e., in the notation of the Hamiltonian in Eq. (S138) those bonds connecting sites within a $\mathcal{D}$ have interaction $J_{pq}$, while dashed lines have interaction $K_{pq}$. Horizontal direction is $\hat{x}$ and vertical is $\hat{y}$. The inter-dimer interactions are different along the $\hat{x}$ and $\hat{y}$ bonds. These are $K'$ and $K$ respectively.

of the field direction - as soon as the defect is polarized by the field in it's direction there should be a non-zero signal. Furthermore, this may also be a possible mechanism for an anomalous thermal-Hall effect.

We may also have a thermal-Hall effect in the case when the spin order is coplanar in the absence of an external magnetic field, provided the above conditions are satisfied. In this case, suppose the spins are coplanar in the $1-3$ plane then an external field in the 2 direction will lead to a non-zero spin component in the 2 direction that is proportional to the field. Thus, this will lead to a thermal-Hall conductivity that is proportional the applied magnetic field.

**B. Presence of spin-orbit coupling.** Let us now consider spin-orbit coupling leading to a DM interaction. Such a term is another route to generate spin-phonon interaction. Recently, spin-phonon interaction mediated by spin-orbit coupling was reported in certain transition metal oxides (5), and the influence of such a term on thermal-Hall has been recently investigated in Ref. (6). We will consider out-of-plane mirror symmetry breaking. This results in the following term,

$$H_{DM} = \sum_{\langle pq \rangle} \vec{D}_{pq} \cdot \left( \vec{S}_p \times \vec{S}_q \right), \qquad [S152]$$

where $\vec{D}_{pq} = D\hat{R}_{pq} \times \hat{z}$ is in the $x-y$ plane, with $\vec{R}_{pq}$ being the vector connecting the sites $p$ and $q$ and $\hat{R}_{pq} = \vec{R}_{pq}/|\vec{R}_{pq}|$. We will restrict ourselves to nearest neighbors only. Let us denote the equilibrium lattice positions by $\vec{R}_p^0$, thus $\vec{R}_{pq}^0$ is the equilibrium value of vector connecting different sites. We will denote the displacement away from these equilibrium positions by $\vec{u}_p$. Let us first expand the DM term,

$$D(\hat{R}_{pq} \times \hat{z}) \cdot (\vec{S}_p \times \vec{S}_q) = D \left[ (\hat{R}_{pq} \cdot \vec{S}_p)(\hat{z} \cdot \vec{S}_q) - (\hat{z} \cdot \vec{S}_p)(\hat{R}_{pq} \cdot \vec{S}_q) \right] \qquad [S153]$$

Let us denote the defect site by index $o$, then the nearest neighbors are at $o + \hat{x}$, $o + \hat{y}$ and $o - \hat{y}$. Now the spin-space and real-space are not independent. The spin-space directions $1, 2, 3$ now align along $x, y, z$ respectively. So, at the

mean-field level the DM term contribution at the defect site is

$$H_{o,DM} = \sum_{i=x,\pm y} D \, \hat{R}_{o,o+\hat{i}} \cdot \left[\vec{S}_o \langle S^3_{o+\hat{i}}\rangle - S^3_o \langle \vec{S}_{o+\hat{i}}\rangle\right] \tag{S154}$$

$$= \sum_{i=x,\pm y} \frac{D}{|\vec{R}_{o,o+\hat{i}}|} \left(\vec{R}^0_{o,o+\hat{i}} + \vec{u}_o - \vec{u}_{o+\hat{i}}\right) \cdot \left[\vec{S}_o \langle S^3_{o+\hat{i}}\rangle - S^3_o \langle \vec{S}_{o+\hat{i}}\rangle\right]. \tag{S155}$$

The DM coupling $D$ is a function of the bond distances and direction. We can expand it as follows:

$$\frac{D}{|\vec{R}_{o,o+\hat{i}}|} = \frac{D}{|\vec{R}^0_{o,o+\hat{i}}|} - \left[1 - \frac{\partial D}{\partial R}\frac{1}{(D/R)}\right] \frac{D}{R}\frac{1}{R}\bigg|_{|\vec{R}^0_{o,o+\hat{i}}|} \hat{R}^0_{o,o+\hat{i}} \cdot (\vec{u}_o - \vec{u}_{o+\hat{i}}),$$

$$\equiv \frac{D}{|\vec{R}^0_{o,o+\hat{i}}|} \left[1 - \frac{\gamma}{|\vec{R}^0_{o,o+\hat{i}}|} \hat{R}^0_{o,o+\hat{i}} \cdot (\vec{u}_o - \vec{u}_{o+\hat{i}})\right]. \tag{S156}$$

where we used the notation, $R \equiv |\vec{R}|$. We will substitute this in the $H_{o,DM}$ and collect all terms to linear order in $u$. We therefore have

$$H_{o,DM} = \sum_{i=x,\pm y} \frac{D}{|\vec{R}^0_{o,o+\hat{i}}|} \bigg[ (\vec{u}_o - \vec{u}_{o+\hat{i}}) \cdot \left[\vec{S}_o \langle S^3_{o+\hat{i}}\rangle - S^3_o \langle \vec{S}_{o+\hat{i}}\rangle\right]$$

$$- \gamma \left[\hat{R}^0_{o,o+\hat{i}} \cdot (\vec{u}_o - \vec{u}_{o+\hat{i}})\right] \hat{R}^0_{o,o+\hat{i}} \cdot \left[\vec{S}_o \langle S^3_{o+\hat{i}}\rangle - S^3_o \langle \vec{S}_{o+\hat{i}}\rangle\right] \bigg] \tag{S157}$$

Thus we have,

$$H_{o,DM} = \frac{D}{R^0} \bigg[ (u^x_o - u^x_{o+\hat{x}})(\langle S^3_{o+\hat{x}}\rangle S^1_o - S^3_o \langle S^1_{o+\hat{x}}\rangle) + (u^y_o - u^y_{o+\hat{x}})(\langle S^3_{o+\hat{x}}\rangle S^2_o - S^3_o \langle S^2_{o+\hat{x}}\rangle)$$

$$+ (u^x_o - u^x_{o+\hat{y}})(\langle S^3_{o+\hat{y}}\rangle S^1_o - S^3_o \langle S^1_{o+\hat{y}}\rangle) + (u^y_o - u^y_{o+\hat{y}})(\langle S^3_{o+\hat{y}}\rangle S^2_o - S^3_o \langle S^2_{o+\hat{y}}\rangle)$$

$$+ (u^x_o - u^x_{o-\hat{y}})(\langle S^3_{o-\hat{y}}\rangle S^1_o - S^3_o \langle S^1_{o-\hat{y}}\rangle) + (u^y_o - u^y_{o-\hat{y}})(\langle S^3_{o-\hat{y}}\rangle S^2_o - S^3_o \langle S^2_{o-\hat{y}}\rangle) \bigg]$$

$$- \frac{\gamma D}{R^0} \bigg[ (u^x_o - u^x_{o+\hat{x}})(\langle S^3_{o+\hat{x}}\rangle S^1_o - S^3_o \langle S^1_{o+\hat{x}}\rangle)$$

$$+ (u^y_o - u^y_{o+\hat{y}})(\langle S^3_{o+\hat{y}}\rangle S^2_o - S^3_o \langle S^2_{o+\hat{y}}\rangle)$$

$$+ (u^y_o - u^y_{o-\hat{y}})(\langle S^3_{o-\hat{y}}\rangle S^2_o - S^3_o \langle S^2_{o-\hat{y}}\rangle) \bigg], \tag{S158}$$

where we used the fact that $\hat{R}^0_{o,o+\hat{x}} = \hat{x}$, $\hat{R}^0_{o,o\pm\hat{y}} = \pm\hat{y}$, with the lattice constant being set to unity, $a = 1$. This also means $R^0 = 1$.

In the continuum limit, $u^j_o - u^j_{o\pm\hat{i}} = \pm\partial_i u^j$, with $i,j = x,y,z$. Using this notation we have,

$$H_{o,DM} = D\bigg[\partial_x u^x S^1_o \langle S^3_{o+\hat{x}}\rangle (1-\gamma) - \partial_x u^x S^3_o \langle S^1_{o+\hat{x}}\rangle (1-\gamma)$$

$$+ \partial_x u^y S^2_o \langle S^3_{o+\hat{x}}\rangle - \partial_x u^y S^3_o \langle S^2_{o+\hat{x}}\rangle$$

$$+ \partial_y u^y S^2_o (\langle S^3_{o+\hat{y}}\rangle - \langle S^3_{o-\hat{y}}\rangle)(1-\gamma) - \partial_y u^y S^3_o (\langle S^2_{o+\hat{y}}\rangle - \langle S^2_{o-\hat{y}}\rangle)(1-\gamma)$$

$$+ \partial_y u^x S^1_o (\langle S^3_{o+\hat{y}}\rangle - \langle S^3_{o-\hat{y}}\rangle) - \partial_y u^x S^3_o (\langle S^1_{o+\hat{y}}\rangle - \langle S^1_{o-\hat{y}}\rangle)\bigg]. \tag{S159}$$

Further we identify the coefficient of the term $\partial_i u^j S_o^\alpha$ by $K_{ij\alpha}$. So from the above Hamiltonian it is clear that $K_{xx1} = D(1-\gamma)\langle S_{o+\hat{x}}^3\rangle$, $K_{xx3} = -D(1-\gamma)\langle S_{o+\hat{x}}^1\rangle$, $K_{xy2} = D\langle S_{o+\hat{x}}^3\rangle$, $K_{xy3} = -D\langle S_{o+\hat{x}}^2\rangle$ and so on. As a simple case, if we consider an in-plane Neel ordering with canting in $\hat{z}$ direction then $\langle \vec{S}_{o+\hat{y}}\rangle = \langle \vec{S}_{o-\hat{y}}\rangle$, then the last two lines in Eq. (S159) vanish. Even in this simple case, the thermal-Hall conductivity is non-vanishing. This may also be a possible mechanism for anomalous thermal-Hall effect.

### C. Spin-Orbit Coupling or Non-coplanar Order.

In this section, we argue for the necessity of spin-orbit coupling or non-coplanar spin order in model B.

Suppose there is no spin-orbit coupling, the microscopic Hamiltonian for the defect must be SU(2) invariant in the absence of external magnetic field, which can be written as

$$H_{micro} = \sum_p J_{op} \vec{S}_o \cdot \vec{S}_p. \qquad [\text{S160}]$$

Here the defect sits at site $o$ and the summation is over neighboring lattice sites $p$. Introducing phonons by modifying the exchange interaction, and assume lattice translation symmetry is preserved, we have

$$H_{micro+ph} = \sum_p J_{op}^{ij} \partial_i u_o^j \vec{S}_o \cdot \vec{S}_p, \qquad [\text{S161}]$$

In the mean-field theory, we can replace treat neighboring spins $\vec{S}_p$ as frozen and replace them by the average $\langle \vec{S}_p\rangle$, so we have

$$H_{def} = \vec{S}_o \cdot \sum_p J_{op} \langle \vec{S}_p\rangle, \qquad [\text{S162}]$$

and

$$H_{ph-def} = \partial_i u_o^j \vec{S}_o \cdot \sum_p J_{op}^{ij} \langle \vec{S}_p\rangle. \qquad [\text{S163}]$$

The information we learn from Eq. (S135) is that we need one coupling to polarize the defect spin, and two other couplings to couple the remaining two perpendicular components to phonons. From Eq. (S162) and Eq. (S163), all these couplings involve some linear combinations of the neighboring spin orders. When the spin order is collinear, we can always choose a frame (this is only allowed without SOC) to make the couplings align in one direction zero, and therefore the thermal Hall effect vanishes. Non-coplanar spin order is a sufficient condition for nonzero thermal Hall effect as there is no frame to make couplings to all three spin directions vanish. There is no definite answer for the case of coplanar magnetic order as it depends on the lattice configuration.

In summary, non-coplanar spin order or spin-orbit coupling are sufficient to give rise to a thermal Hall effect.

### 10. Order-of-Magnitude Estimate for Model B

In this part we present an order-of-magnitude estimate of the ratio $|\kappa_H/\kappa_{xx}|$ within model $B$.

Assuming that there are $N_i$ defects in the system, our result for model $B$ reads

$$\kappa_H^{sj} = \frac{N_i}{N_{\text{sys}}} \frac{\beta^2 \Delta^4 \text{csch}(\beta\Delta)}{30\pi \Gamma_{\text{ph}} m} \left( c_L^{-3} K_L + c_T^{-3} K_T \right). \qquad [\text{S164}]$$

By dimensional analysis, $K_{ij\alpha}$ has the dimension of energy. Microscopically, it comes from expanding the exchange coupling $J(a)$ as a function of inter-atomic distance $a$: $K_{ij\alpha} \sim a\partial_a J(a)$. Assuming, $\partial_a J(a) \simeq J(a)/a$, we conclude that $K_{ij\alpha} \sim J(a)$. Notice that the local splitting the spin defect also comes from exchange, we therefore assume $K_{ij\alpha} \sim \Delta$.

24

To have a linear-in-field thermal Hall effect, we consider a scenario where coplanar spin-order at zero field is canted by external field. We choose the spin frame such that when external field $B = 0$, the couplings are such that $K_{ij1} \sim \Delta$, $K_{ij2} = 0$. Due to canting of the spins, we would expect $K_{ij2} \sim \Delta \theta_B$, with the canting angle $\theta_B \sim \Delta_B/\Delta \ll 1$. Here $\Delta_B \simeq 2\mu_B B$ is the splitting due to external magnetic field. The quadratic combinations of couplings therefore scales as $K_L, K_T \sim \Delta \Delta_B$. So the thermal Hall effect scales as

$$\kappa_H^{sj} = \frac{N_i}{N_{sys}} \frac{\Delta^5 \Delta_B}{T^2 \sinh\left(\frac{\Delta}{k_B T}\right)} \frac{1}{m\Gamma_{\text{ph}} c^3} \frac{A_H}{k_B \hbar^3} . \qquad [S165]$$

Here $A_H$ is a numerical coefficient depending on the detailed expressions of $K_L$ and $K_T$, $c$ is some averaged sound velocity and we have restored $k_B$ and $\hbar$.

We compare this with the simple kinetic model for longitudinal thermal conductivity

$$\kappa_{xx} \simeq \frac{1}{3} \frac{c^2}{\Gamma_{\text{ph}}} C_V , \qquad [S166]$$

where the heat capacity $C_V$ follows the Debye model

$$C_V = \frac{12\pi^4}{5} \left(\frac{T}{T_D}\right)^3 k_B \frac{\rho}{m} . \qquad [S167]$$

Here $\rho$ is the mass density and $m$ is the ion mass and $T_D$ is the Debye temperature.

The ratio is therefore

$$\frac{\kappa_H}{\kappa_{xx}} = \frac{5\Delta_B k_B^3 T_D^3}{4\pi^4 c^5 \rho \hbar^3} A_H \frac{N_i}{N_{sys}} \Phi\left(\frac{\Delta}{k_B T}\right), \quad \Phi(x) = \frac{x^5}{\sinh(x)} . \qquad [S168]$$

Plugging in values of $\rho = 6000 \text{kg/m}^3$, $c = 5000\text{m/s}$, $T_D = 400\text{K}$, $\Delta_B = 2\mu_B B$ where $B = 15\text{T}$, we have

$$\frac{\kappa_H}{\kappa_{xx}} = 2.7 \times 10^{-5} A_H \frac{N_i}{N_{sys}} \Phi\left(\frac{\Delta}{k_B T}\right) . \qquad [S169]$$

Remarkably, the maximum of this function is independent of value of $\Delta$, which is also a measure of the phonon-spin coupling. The function $\Phi$ is peaked at $k_B T \simeq 0.2\Delta$, where we have

$$\frac{\kappa_H}{\kappa_{xx}} = 1.2 \times 10^{-3} A_H \frac{N_i}{N_{sys}} . \qquad [S170]$$

Experimentally, the peak occurs at around 20K, and therefore $\Delta$ is about 100K.

If the thermal Hall effect in cuprate is due to phonons scattering on spin glass, essentially every spin participates in phonon scattering and we can set $N_i/N_{sys} = 1$. Then the ratio is indeed close to the experiment, provided that $A_H$ is around order one.

# Part III

# Intrinsic contribution in Quadratic System

### 11. Berry Curvature Formula for Intrinsic Contribution

To demonstrate the lattice formalism, we show that it agree with the Berry curvature formula for phonon Hall effect derived in Ref. (7). We assume there is only a quadratic phonon system with no dissipation $\Gamma_{\text{ph}} = 0$, and there is no defect present.

**A. Kubo contribution to Thermal Hall Conductance.** The Kubo contribution is the same as Eq. (S71), we reproduce it here without antisymmetrizing $f, g$:

$$\kappa_{xy}^{\text{Kubo}}(f,g) = \frac{\beta}{16\pi} \int \mathrm{d}z\, n_B(z) \operatorname{Tr} \big[ ([hJh, f])(-D'_+)([hJh, g])(D_+ - D_-) \\ - ([hJh, f])(D_+ - D_-)([hJh, g])(-D'_-) \big] \quad [\text{S171}]$$

**B. Magnetization Correction.** We now consider arbitrary variation of the quadratic Hamiltonian. The expression for magnetization correction is again given by

$$\mu^E(\delta f \cup \delta g) = -\frac{\beta}{4} \big[ 2 \langle\langle dH(g); J^E(\delta f) \rangle\rangle - 2 \langle\langle dH(f); J^E(\delta g) \rangle\rangle + \langle\langle dH; f J^E g - g J^E f \rangle\rangle \big], \quad [\text{S172}]$$

where

$$dH(f) = \sum_p f_p dH_p = \frac{1}{4}\zeta(f\mathrm{d}h + \mathrm{d}h f)\zeta, \quad [\text{S173}]$$

$$dH = \sum_p dH_p = \frac{1}{2}\zeta \mathrm{d}h \zeta. \quad [\text{S174}]$$

The formula we need is

$$-\beta \langle\langle \zeta A \zeta; \zeta B \zeta \rangle\rangle \\ = -2 \int \frac{\mathrm{d}z}{2\pi i} n_B(z) \operatorname{Tr}[AD_+ B(D_+ - D_-) + A(D_+ - D_-)BD_-]. \quad [\text{S175}]$$

We obtain

$$\mu^E(\delta f \cup \delta g) = \frac{-1}{32\pi i} \int \mathrm{d}z\, n_B(z) \operatorname{Tr} \big[ D_+ \mathrm{d}h D_+ J^{-1}([[Jh, f], [Jh, g]]) \\ + [(Jh)^2, f] D_+ J^{-1}[Jh, g] - [(Jh)^2, g] D_+ J^{-1}[Jh, f] \\ - [Jh, g] D_+ J^{-1}[(Jh)^2, f] + [Jh, f] D_+ J^{-1}[(Jh)^2, g] \big) - (D_+ \to D_-) \big]. \quad [\text{S176}]$$

**C. Berry Curvature formula.** We now follow Ref. (1), choose a scaling variation $\mathrm{d}h = h$ which will yield derivative of $\kappa_{xy}$ with respect to temperature. By inspection, the algebraic form of Eq. (S171) and Eq. (S176) are exactly $1/2$ times (F18) and (F19) of Ref. (1), given that we substitute $G_\pm \to D_\pm(iJ)^{-1}$, $h \to iJh$, $\mathfrak{f} \to n_B$. Therefore, following the same algebra that lead to (F23) in Ref. (1) we conclude that

$$\frac{\mathrm{d}}{\mathrm{d}T}\left(\frac{\kappa_{xy}(f,g)}{T}\right) = -\frac{i}{4\pi T^3} \int_{-\infty}^{\infty} \mathrm{d}z\, n'_B(z) \operatorname{Tr}\big[[h,f]D_+ J^{-1} D_+[h,g] z^3 (D_+ - D_-) \\ - [h,g]D_- J^{-1} D_-[h,f] z^3 (D_+ - D_-) \big] \\ = \frac{1}{4\pi T^3} \int_{-\infty}^{\infty} \mathrm{d}z\, n'_B(z) \operatorname{Tr}\big[[iJh,f](-iD_+ J^{-1})^2 [iJh,g] z^3 (D_+ - D_-)(iJ)^{-1} \\ - [iJh,g](-iD_- J^{-1})^2 [iJh,f] z^3 (D_+ - D_-)(iJ)^{-1} \big]. \quad [\text{S177}]$$

We can now evaluate Eq. (S177) in the continuum limit. We set $f = -x/L$ (the average of $\theta$ functions) and $g = -y/L$. In the continuum limit we therefore have

$$[iJh, f] = \frac{i}{L} \frac{\partial iJh(k)}{\partial k_x}, \quad [iJh, g] = \frac{i}{L} \frac{\partial iJh(k)}{\partial k_y}. \quad [\text{S178}]$$

Switching to the basis where $iJh$ is diagonal, we have

$$iJh(k) = M(k)\mathcal{E}(k)M(k)^{-1}. \quad [\text{S179}]$$

26

The Green's functions are

$$D_{\pm}(z,k)(iJ)^{-1} = \frac{1}{z \pm i0 - iJh(k)} = M(k)\frac{1}{z \pm i0 - \mathcal{E}(k)}M(k)^{-1}. \qquad [S180]$$

The trace factorizes into momentum sum and band trace:

$$\text{Tr}\, X = L^d \int \frac{\mathrm{d}^d k}{(2\pi)^d} \sum_i X_{ii}(k). \qquad [S181]$$

The thermal conductivity is therefore

$$\frac{\mathrm{d}(\kappa_{xy}/T)}{\mathrm{d}T} = \frac{1}{2T^3} \int \frac{\mathrm{d}^d k}{(2\pi)^d} \sum_i n'_B(\mathcal{E}_{ik})\mathcal{E}_{ik}^3 \Omega_{ik}^z. \qquad [S182]$$

Here the sum is over both positive and negative energy modes, both of which contribute equally.

The Berry curvature is defined as

$$\Omega_{ki}^z = -i[A_x, A_y]_{ii}, \qquad [S183]$$

where the connection is

$$A_{\mu,ij} = \left(M^{-1}\frac{\partial M}{\partial k^\mu}\right)_{ij}. \qquad [S184]$$

It can be checked that Eq. (S182) agrees with Ref. (7).



\end{document}